\documentclass[11pt,reqno,letterpaper]{amsart}

\usepackage{amsmath}
\usepackage{amssymb}
\usepackage{graphicx}
\usepackage{xr}

\def\R{\mathbb R}
\def\Z{\mathbb Z}
\def\N{\mathbb N}

\def\r{\rangle}
\def\l{\langle}

\def\define{\overset{\text{\tiny def}}{=}}

\def\O{\mathop{\operatorname{{O}}}\nolimits}
\def\U{\mathop{\operatorname{{U}}}\nolimits}
\def\SU{\mathop{\operatorname{{SU}}}\nolimits}
\def\Sp{\mathop{\operatorname{{Sp}}}\nolimits}

\def\G{\mathop{\operatorname{{G}}}\nolimits}
\def\E{\mathop{\operatorname{{E}}}\nolimits}

\def\Cf{\mathop{\operatorname{{C}}}\nolimits}
\def\Sf{\mathop{\operatorname{{S}}}\nolimits}

\def\Aone{\mathop{\operatorname{{A_1}}}\nolimits}
\def\Atwo{\mathop{\operatorname{{A_2}}}\nolimits}

\def\Ctwo{\mathop{\operatorname{{C_2}}}\nolimits}
\def\Gtwo{\mathop{\operatorname{{G_2}}}\nolimits}
\def\aff{\mathop{\operatorname{{aff}}}\nolimits}

\numberwithin{equation}{section}

\begin{document}

\title[Two dimensional exponential transforms]
{Discrete and continuous exponential transforms\\
 of simple Lie groups of rank two}
\author{I.~Kashuba$^\dag$, J.~Patera$^\ddag$}
\address{$^\dag$Departamento de Matemática,
Instituto de Matemática e Estatística da Universidade de São Paulo,
Rua do Matão, 1010 - Cidade Universitária
CEP 05508-090, São Paulo - SP - Brasil}

\email{kashuba@ime.usp.br}

\address{$^\ddag$Centre de recherches  math\'ematiques, Universit\'e
de Montr\'eal, C.P. 6128 succ. Centre-Ville, Montr\'eal, Qu\'ebec
H3C\,3J7, Canada.}

\email{patera@crm.umontreal.ca}

\begin{abstract}
We develop and describe continuous and discrete transforms of
class functions on compact simple Lie group $G$ as their
expansions into series of uncommon special functions, called here
$\,\E$-functions in recognition of the fact that the functions
generalize common exponential functions. The rank of $\,G\,$ is
the number of variables in the $\,\E$-functions. A uniform
discretization of the decomposition problem is described on
lattices of any density and symmetry admissible for the Lie group $G$.
\end{abstract}

\subjclass{33E99; 42B99; 42C15; 20F55}

\maketitle

\section{Introduction}\label{IntroSection}

The aim of this paper is to generalize common exponential functions in one variable $x$,
\begin{equation}\label{efunc}
\E_m(x):=e^{imx}\,,\quad m\in\Z\,,\ x\in\R\,,
\end{equation}
together with the corresponding Fourier transform,
\begin{equation}\label{1Dtransform}
f(x)=\sum_{m=0}^\infty b_m e^{imx}\,,\qquad
b_m=\frac1{2\pi}\int_0^{2\pi}f(x)e^{-imx}\,dx\,,
\end{equation}
to any number of variables.

For close to two centuries functions \eqref{efunc} have been part
of Fourier analysis. Crucial property is their pairwise
orthogonality when integrated over a range  $a\leq x\leq a+2\pi$
with any $a\in\R$. More recent but equivalent interpretation of
the functions is as irreducible characters of the 1-parametric
unitary group $\U(1)$. There is yet another interpretation of the
functions \eqref{efunc}, rather trivial in this case, which
nevertheless is the departure point for our generalization. It is
presented in the example of Section~3.

The simplest possible $n$-dimensional generalization of
exponential functions, is based on the $n$-fold product
$\U(1)\times \U(1)\times\cdots\times \U(1)$. Corresponding
functions are products of $n$ copies of $\E_m(x)$, each depending
on its own continuous variable $x$ and its own lattice variable
$m$. It is undoubtedly useful and it is frequently used. However
we are not concerned with such a generalization in this paper.

Underlying symmetry group of $\E$-functions in this article is any
compact semisimple Lie group $G$ of rank $n$ in general. The rank
of $G$ is the number of continuous variables. Here our
considerations are focused on the three simple Lie groups of rank
2, namely $\SU(3)$, $\O(5)$ (or $\Sp(4))$, and $\G(2)$. Our aim is
to describe the three cases in a ready-to-use form.

There are two papers dealing briefly with $\E$-functions: Their
definition appeared in \cite{P}, and their orthogonality is proven
in \cite{MP1}. We know of no other attempt in the literature to
generalize exponential functions to more than one variable.

First however let us underline a close relation of the
$\E$-functions of this paper with the $\Cf$- and $\Sf$-functions
of \cite{PZ1,PZ2,PZ3,KP,KP2}. All three families of functions are
based on a semisimple compact Lie group $G$. They are constructed
by summation of appropriate products of $\U(1)$ characters over an
orbit of a relevant finite group. It is the Weyl group $W$ of $G$
in case of $\Cf$- and $\Sf$-functions, and it is the even subgroup
$W_e\subset W$ of $G$ in case of the $\E$-functions. Furthermore,
due to the relation of the three families to the group elements of
the maximal torus of the underlying Lie group, one gets the
symmetries of the functions with respect to both affine Weyl group
$W^{\aff}$ and even affine Weyl group $\,W_e^{\aff}.$

For comparison with \eqref{efunc}, let us point out that
1-dimensional $\Cf$- and $\Sf$-functions are (up to a
normalization) the familiar trigonometric functions
$$
\Cf_m(x)\sim\cos mx,\quad \Sf_m(x)\sim\sin mx\,,\quad
    m\in\Z^{\geq0}\,,\ x\in\R\,.
$$
Within either $\Cf$- or $\Sf$-family, the functions are pairwise
orthogonal in $a\leq x\leq a+\pi$. The underlying Lie group is
$\SU(2)$ for both families.

Recently $\Cf$-functions were studied relatively extensively.
Their properties are reviewed in \cite{KP} (see also references
therein), similarly the $\Sf$-functions are found in
\cite{P,MP1,KP2}. Let us as well point out that some properties of
functions symmetrized over a finite group in general are described
by I. Macdonald, \cite{IM}.

It may appear (falsely) that the expansions, based on compact
semisimple Lie groups, impose severely constraining requirements
on functions amenable to such expansions. Typically one is
interested in expansions of a function given on a finite region,
say $R$, of a real Euclidean space $\R^n$ either as a continuous
function  (`continuous data') or by its values at lattice points
in $R$  (`digital data').  In our approach one first needs to
choose a semisimple Lie group of rank $n$ whose weight lattice has
the same geometric structure as the lattice of the data  (there is
always at least one such group).  Then the region $R$ is inserted
into the fundamental region $F$ of the chosen Lie group $R\subset
F$. In the  case of digital data, a unique aspect of our method is
the easy possibility to match the density of the data points in
$R$ by the density of a grid $F_M$ in $F$. More precisely, the
positive integer $M$ selects a finite Abelian subgroup of the Lie
group such that its conjugacy classes are represented by the
points of $F_M$.

The families of $\Cf$-, $\Sf$-, and $\E$-functions have a number
of properties in common. In addition to their orthogonality, when
integrated over a finite region $F$ of an Euclidean space $\R^n$,
they are also discretely orthogonal when summed up over a discrete
grid $F_M\subset F$. Discrete orthogonality of $\Cf$-functions in
general is the main content of \cite{MP2}. Furthermore, the
lattice, obtained when $F_M$ is extended to the whole space
$\R^n$, is setup uniformly for all three families. Density of such
a lattice is specified by a positive integer $M$.  Functions of
the three families are eigenfunctions of the same Laplace
operator, namely the one appropriate for the group $G$, differing
mainly by their behavior at the boundary of $F$. Their eigenvalues
are known explicitly for all $n$ and all three families. Their
products are decomposable into their sums. The functions can be
built up recursively (in lattice variables), for any number of variables
$x\in\R^n$, by a judicious choice of the lowest few  and by their
successive multiplication. In principle they could be also built
recursively in the points of $F_M$ (for the same lattice variable), using
the fact that the points of $F_M$ stand for conjugacy classes of a finite
Abelian group, although it could be a laborious way to do it.

Discrete orthogonality of $\Cf$-functions, see \cite{MP2}, were
exploited in challenging mathematical applications, see \cite{GP}
and references therein. Immediate motivation of our current
interest in the three families of functions arose as a result of
the observation made in \cite{Wa,Ag,AP1} that continuous
extensions of the (finite) discrete expansions of functions on
$F_M$ smoothly interpolate digital data between  discrete points.
Such an observation is strongly supported by numerous convincing
examples and qualitative arguments in  case of $\Cf$-functions and
certainly carries over to $\Sf$-functions. However, a quantitative
demonstration has yet to be made even in those cases.

In two dimensions practical need to interpolate digital data lead
to development of a number of sophisticated interpolation methods.
Comparison of such methods with ours depends on the model
functions one compares it with. In general, one may say that the
precision of the best interpolation methods is comparable to ours.
However, unlike our approach, none of these methods readily
generalizes to higher dimensions. In our case all one needs is to
replace one compact semisimple Lie group of rank two by another one of
rank $\,n<\infty$.

Generalization of the $\E$-transform from one to more dimensions
in case of a semisimple Lie group which is not simple, say
$G=G_1\times G_2$, presents two interesting options, each worth to
be explored. The fundamental region $F^e$, where the expansion
takes place, and the expansion functions are different. In spite
of that one has in both cases the continuous and discrete
orthogonality of the functions in $F^e$. The simpler option of the
two is followed up in this paper.

In Section 2 we recall briefly the definition of the Weyl group of
simple (or semisimple) Lie group and its affine Weyl group and
their basic properties. Also we define the subgroup of even
elements of the Weyl group $\,W_e\,$ and the corresponding even
affine Weyl group together with their root lattice, fundamental
region, etc. In Section~3 we introduce $\,\E$-functions
$\E_\lambda(x)$ of a Weyl group. The functions are specified by a
given point $\lambda\in\R^n$. Their $\,W_e\,$-invariance is shown.
Also, analogously to the case of both $\,\Cf$- and
$\,\Sf$-functions, $\,\E$-functions are orthogonal over the
fundamental region $\,F^e\subset \R^n$ of $\,W_e$. The general
method of expansion of a function on $\,F\,$ into the sum of
orthogonal $\,\E$-functions is given. We illustrate it on the case
of the rank one Lie group $\,\Aone$. In Section~4 the
$\,\E$-functions together with their continuous transforms are
described for the three simple Lie groups of rank $\,2$. Discrete
orthogonality of $\,\E$-functions in general is the content of
Section~5, while in Section~6 pertinent properties are described
for exploitation of continuous extensions of discrete
$\E$-expansions of functions on the fundamental region $\,F^e\,$
for the simple Lie groups of rank two. The decomposition of the
product of $\,\E$-functions for these groups is the subject of
Section~7. Finally, in Section~8 we introduce central splitting of
functions given on $F$ or $F_M$ into the sum of $\,s\,$ functions,
where $\,s\,$ is the order of the center of the corresponding Lie
group. Each component function has simpler $\,\E$-functions
expansions. Concluding remarks and some related problems are
brought forward in Section~9.

\section{Weyl group, its even subgroup and their affininizations}
\label{general}

Let $\,r_i\,$ be reflection transformation of $\,\R^n\,$ with
respect to $\,n-1$-dimensional subspace containing the origin.
Consider finite groups $\,W\,$ generated by $\,n\,$ such
reflections $\,r_1,r_2,\dots, r_n$. For any point $\,\lambda\in
\R^n\,$ we define the orbit $\,W(\lambda)\,$ of the point
$\,\lambda\,$ under the action of $\,W\,$ as the set of all
different points of the form $\,\omega\lambda,$ $\,\omega\in W$.
Then the corresponding orbit function is the following
\begin{equation}\label{orbit_function}
\Cf_{\lambda}(x)=\sum_{\mu\in \,W(\lambda)} e^{2\pi i\langle
\,\mu, x\rangle}\,, \qquad x\in \R^n,
\end{equation}
where $\,\langle\ ,\ \rangle\,$ is a scalar product in $\,\R^n$.
Note that for $\,n=1\,$ we have $\,\Cf_{\lambda}(x)=2\cos(\pi
mx)$, where $\,m\in \Z^{\geq0}\,$ and $\,x\in\R$.

In this paper we consider $\,\E$-functions which are orbit
functions corresponding to symmetry group $\,W_e\,$ of even
elements of Weyl groups of simple (or semisimple) Lie group. Below
we recall some basic definition about both Lie groups and
corresponding Weyl groups. For further information about both
simple Lie groups and the Weyl groups we refer to the books
\cite{Hum}, \cite{Kane}.

The Weyl group $\,W\,$ of any  simple (or semisimple) Lie group is
specified by its Coxeter-Dynkin diagrams. The diagram is a concise
way to give a certain non-orthogonal basis
$\,\Pi=\{\alpha_1,\dots,\alpha_n\}\,$ in $\,\R^n$. Each node of
the diagram is associated with a basis vector $\,\alpha_k\,$,
called the simple root of the Lie group. Acting by elements of the
Weyl group $\,W\,$ upon simple roots, we obtain a finite system of
vectors, which is invariant with respect to $\,W$. A set of all
such vectors is called the root system $\,\Delta\,$ associated
with a given Coxeter-Dynkin diagram. The set of all linear
combinations
\begin{equation}\label{q_lattice}
Q=\left\{\sum_{i=1}^n a_i\alpha_i\ |\,
a_i\in\Z\right\}=\bigoplus_{i=1}^n\Z\alpha_i
\end{equation}
is called the root lattice. Relative lengths and angles between
simple roots of the basis $\,\Pi\,$ are specified in terms of the
elements of the Cartan matrix $\,C=(c_{ij})_{i,j=1}^n\,,$ where
$$
c_{ij}=\tfrac{2\langle\alpha_i|\alpha_j\rangle}{\langle\alpha_j|\alpha_j\rangle}=
\langle\alpha_i|\check{\alpha}_j\rangle, \quad {\rm for} \quad
i,j=1,\dots,n.
$$
Here $\,\check{\alpha}_j\,$ is the simple root of the dual root
system
$\,\check{\Delta}=\{\check{\alpha}=2\alpha/\langle\alpha,\alpha\rangle\
|\ \alpha\in\Delta\}$. Denote by $\,\check{Q}\,$ the corresponding
coroot lattice. Absolute length for the roots is chosen by an
additional convention, namely that the longer roots of $\,\Pi\,$
satisfy $\,\langle\alpha|\alpha\rangle=2$. In addition to the
$\,\alpha$-basis it is convenient to introduce the basis of
fundamental weights $\,\omega_1,\dots,\omega_n$. The
$\,\omega$-basis and $\,\alpha$-basis are related by the inverse
of the Cartan matrix
$$
\omega_j=\sum_{k=1}^n (C^{-1})_{jk}\alpha_k.
$$
Analogously to the root lattice we introduce the weight lattice
$$
P=\left\{\sum_{i=1}^n a_i\omega_i\ |\,
a_i\in\Z\right\}=\bigoplus_{i=1}^n\Z\omega_i\,.
$$
We also define the set of dominant weights $\,P^+\,$ and the set
of strictly dominant weights $\,P^{++}\,$
\begin{equation}\label{P^+}
P^+=\Z^{\geq 0}\omega_1+\Z^{\geq 0}\omega_2+\dots+\Z^{\geq
0}\omega_n \  \supset \  P^{++}=\Z^{> 0}\omega_1+\Z^{>
0}\omega_1+\dots+\Z^{> 0}\omega_1.
\end{equation}

For each $\,\alpha\in\Delta\,$ and integer $\,k\,$ we define the
hyperplane
$$
H_{\alpha,\,k}=\{t\in \R_n\,|\,\langle t,\alpha\rangle=k\}
$$
and the associated  reflection $\,r_{\alpha,\,k},$ in the
hyperplane $\,H_{\alpha,\,k}\,$
\begin{equation}\label{affine_reflection}
r_{\alpha,\,k}\,x=x-\langle \alpha,\,
x\rangle\check{\alpha}+k\check{\alpha}.
\end{equation}
The finite Weyl group $\,W\,$ is generated by $\,r_{\alpha,\,0},$
$\,\alpha\in\Pi$. Since the action of $\,W\,$ on $\,\Pi\,$ gives
the root system $\,W\Pi=\Delta$, $\,W\,$ can be extended to the
affine Weyl group $\,W^{\aff}\,$, the group generated by
$\,r_{\alpha,\,k}\,$ for all $\,\alpha\in\Pi\,$ and $\,k\in\Z$.
$\,W^{\aff}\,$ is an infinite group such that
\begin{equation}\label{decomposition}
\,W^{\aff} =\check{Q}\rtimes W \end{equation}
It is the semidirect product of its subgroups $\,W\,$ and the invariant Abelian
subgroup $\,\check{Q}\,$, the coroot lattice, for proof see \cite{Kane}.

For any Weyl group there exists a unique highest root
$$\xi_{h}=\sum_{i=1}^n m_i\alpha_i\equiv\sum_{i=1}^n
q_i\check{\alpha}_i.$$ Coefficients $\,m_i\,$ and $\,q_i\,$ are
called marks and comarks correspondingly and could be found in
\cite{BMP}

Finally, for any affine Weyl group we introduce its fundamental
domain (or region) $\,F\subset \R^n\,$ as the convex hull of $\,\{
0,\,\tfrac{\check{\omega}_1}{q_1},\dots,\tfrac{\check{\omega}_n}{q_n},\}$
where $\,q_1,\dots,q_n$. By definition $\,F\,$ is closed. If
$\,G\,$ is not simple then its fundamental region is the Cartesian
product of fundamental regions of its simple components.

\subsection{Subgroup of even elements of the Weyl group $\,W_e\,$ and its affine group $W_e^{\aff}$}\

Let $\,W\,$ be a Weyl group of simple Lie group. This group is
generated by reflection transformations $\,r_i,\,$ $i=1,\dots,n$.
We consider the subset of $\,W\,$
$$
W_e=\langle r_{i_1}\dots r_{i_p}|\ p\ {\rm is\  even},\
i_j\in\{1,\dots,n\}\rangle,
$$
i.e. the set of elements generated by even number of reflections,
or the elements of $\,W\,$ of even length. Obviously, $\,W_e\,$
forms a finite normal subgroup of $\,W$ of index $\,2\,$, such
that
\begin{equation}\label{W_e_w_relation}
W=W_e\dot{\cup} \bigcup_{i=1}^n r_i W_e\equiv W_e\dot{\cup} r_i
W_e, \qquad |W_e|=\tfrac12|W|.
\end{equation}

The index $\,i\,$ in the last equation is arbitrary, since  $\,r_j
{r_i}^{-1}\in W_e\,$ for any $i,j\in\{1,\dots,n\}$. For an
arbitrary point $\,\lambda\,$ in $\,\R^n$ we denote by
$\,W_e(\lambda)\,$ its orbit with respect to the action of the
even Weyl group. Every $\,\lambda\,$ is contained in precisely one
$\,W_e$-orbit. From \cite{KP} it follows that each original
$\,W$-orbits contains a unique $\,\mu\in P^{+}$. By
\eqref{W_e_w_relation} we obtain that each $\,W_e$-orbit contains
a unique element belonging to $\,P_e:=P^+\cup r_i P^{++}$. The
$\,W$- and $\,W_e$-orbits are in the following correspondence with
the orbits of original Weyl groups:
\begin{equation}\label{rel_orb}
W(\lambda)=
\begin{cases}
W_e(\lambda)\cup W_e(r_i\lambda),    &\text{if $\,\lambda\in
P^{++}$}
\quad {\rm for\ some \ } \ i\in\{1,\dots,n\}\\
W_e(\lambda), &\text{if $\,\lambda \in P^{+}\setminus P^{++}.$}
\end{cases}
\end{equation}

In particular, let we denote by $\,|W_e(\lambda)|\,$ the size of
the orbit $\,W_e(\lambda)\,$, then $\,|W_e(\lambda)|\,$ is either
equal to $\,|W(\lambda)|\,$ or to $\,\tfrac12|W(\lambda)|\,$.

Consider the original affine group $\,W^{\aff}\,$ generated by
$\,r_{\alpha,\,k}\,$ defined in \eqref{affine_reflection}. Then
the even affine group is the subgroup of words of even length in
$\,r_{\alpha,\,k}\,$ of $\,W^{\aff}\,$, i.e.
\begin{equation}
W_e^{\aff}=\langle r_{\alpha_{i_1},\,k_1}\dots
r_{\alpha_{i_p},\,k_{p}}|\ p\ {\rm is\ even},\
\alpha_{i_j}\in\Pi,\
 k_i\in\Z\rangle.
\end{equation}
As in the case of original affine Weyl group we have the following
relation between $\,W_e^{\aff}\,$ and $\,W_e\,$
\begin{equation}\label{decop_th}
W_e^{\aff}=\check{Q}\rtimes W_e. \end{equation} Indeed, the
subgroup of $\,W_e^{\aff}\,$ generated by $\,r_{\alpha,\,0}\,$
coincides with $\,W_e$, therefore $\,W_e < W_e^{\aff}$. For any
element $\,d\in \R^n\,$ define a translation
$$
\begin{array}{c}
\tau(d)\,x=x+d, \qquad x\in\R^n
\end{array}
$$
For any two $\,d,d'\in \R^n\,$ $\,\tau(d)\,\tau(d')=\tau(d+d')$,
therefore we may identify $\,\check{Q}\,$ with a group of
translations on $\,\R^n\,$. Since
\begin{equation}\label{refl_trans}
r_{\alpha,\,k}=\tau(k\check{\alpha})r_{\alpha,\,0} \end{equation} we
obtain that for $\,\alpha\in\Pi\,$ and $\,k\in\Z\,$
$\,\tau(k\check{\alpha})=r_{\alpha,\,k} r_{\alpha}^{-1}\in
W_e^{\aff}$. In particular for any $\,i=1,\dots,n\,$
$\,\tau(\check{\alpha_i})\in W_e^{\aff}\,$ and therefore
$\,\check{Q}\in W_e^{\aff}$. Further, from \eqref{refl_trans}
follows that $\,W_e^{\aff}=W_e\,\check{Q}\,$. Also since any
non-zero element of $\,\check{Q}\,$ has infinite order and any
non-zero element of $\,W_e\,$ has finite order we obtain that
$\,W_e\cap\check{Q}={0}$. Finally, we have to show that subgroup
$\,\check{Q}\,$ is normal in $\,W_e^{\aff}$. Indeed, for any $\,w\in
W_e\subset W\,$ and any $\,d\in\check{Q}\,$ we have
$\,w\,\tau(d)w^{-1}=\tau(wd)\,$.

Using \eqref{W_e_w_relation} we also can define now the
fundamental domain of $\,W_e^{\aff}\,$ being a set
\begin{equation}\label{fund_region}
F^e=F\cup
r_i F.
\end{equation}

\section{Definition of $\,\E$-functions, their relations
to $\,\Cf$-functions}

We start with the definition \eqref{orbit_function} of
$\,\Cf$-functions. The $\,\Cf$-function $\Cf_{\lambda}(x)$ is the
contribution to an irreducible character from the orbit
$\,W(\lambda)\,$, $\lambda\in P^+$. If in \eqref{orbit_function}
we restrict ourselves to the orbit $\,W_e(\lambda)\,$ instead of
the orbit of $W$, we obtain the $\,\E$-function $\E_{\lambda}(x)$:
\begin{equation}\label{E-function}
\E_{\lambda}(x)=\sum_{\mu\in \,W_e(\lambda)} e^{2\pi i\langle
\mu,\,x\rangle} \ ,
  \qquad {\rm for\ } x\in \R^n, \quad \lambda\in P.
\end{equation}
The $\,\Cf$-functions appeared in \cite{MP2,KP} under the name
"orbit functions". Their many properties, very useful for
applications, were extensively studied i n \cite{P}-\cite{KP}. In
this section we formulate analogous properties of
$\,\E$-functions.

To start, both families of $\,\Cf$- and $\,\E$-functions are based on semisimple
Lie algebra, the rank of the algebra is the number of
variables. They are given as the finite sums of exponential
functions, therefore they are continuous and have derivatives of
all orders in $\R^n$.

\subsection{$\,W_{e\ }$- and $\,W_e^{\aff}\,$-invariance of $\E$-functions}\

 The $\,\Cf$-functions \eqref{orbit_function} are invariant under the action
 of $\,W$. For any $\,\lambda\in P\,$
$\,\E_{\lambda}(x)$ is invariant under the action of $\,W_e$.
Indeed,
\begin{equation}
\E_{\lambda}(wx)
  =\sum_{\mu\in \,W_e(\lambda)} e^{2\pi i\langle\mu,\,wx\rangle}
  =\sum_{\mu\in W_e(\lambda)} e^{2\pi i\langle w^{-1}\mu,\,x\rangle}
  =\E_{\lambda}(x)
      \quad\text{for any}\quad w\in W_e,
\end{equation}
since the scalar product $\,\langle\ ,\ \rangle\,$ is invariant
with the respect to $\,W\,$ and $\,W_e(\mu)=W_e(w^{-1}\mu)\,$.

The $\,\Cf$-function corresponding to $\,\lambda\in P\,$ is
invariant under the action of $\,W^{\aff}$. Let us show that
$\,\E$-functions for $\,\lambda\in P\,$ are invariant with respect
to $\,W_e^{\aff}$. Since $\,W_e^{\aff}=\check{Q}\rtimes W_e\,$ it
is enough to show invariance of $\,\E_{\lambda}(x)\,$ with respect
to any translation $\,\tau(d)$, $\,d\in\check Q$. For
$\,\lambda\in P\,$ any $\,\mu\in W_e(\lambda)\,$ also belongs to
$\,P\,$ hence
\begin{equation} e^{\langle \mu,\tau(d)x\rangle}=e^{\langle \mu,
d\rangle+\langle \mu,x\rangle}=e^{{\rm
integer}+\langle\mu,x\rangle}=e^{\langle\mu,x\rangle}.
\end{equation}

Since $\,\E$-functions are invariant under the action of
$\,W_e^{\aff}\,$, it is enough to consider them only on the
fundamental domain $\,F^e$. The values on other points of
$\,\R^n\,$ are determined by using the action of $\,W_e^{\aff}\,$
on $\,F^e$.

\subsection{Relation between $\,\E$- and $\,\Cf$-functions}\

The original Weyl group also acts on the $\,\E$-functions. By
\eqref{rel_orb} for any $\,\lambda\in P^+\setminus P^{++}\,$ we
obtain that $\,\E_{\lambda}(r_ix)=\E_{\lambda}(x)$, if
$\,\lambda\in P^{++}\,$ then
$\,\E_{\lambda}(r_ix)=\E_{r_i\lambda}(x)$. Bringing it all
together, we obtain
\begin{equation}\label{c_via_e}
\Cf_{\lambda}(x)=
    \begin{cases}
\E_{\lambda}(x)+\E_{r_i\lambda}(x) &\text{if $\lambda\in P^{++},$}\\
\E_{\lambda}(x), &\text{otherwise.}
    \end{cases}
\end{equation}
We call $\,\lambda\in P_e\,$ an intrinsic point if $\,\lambda\in
P^{++}\cup r_iP^{++}$.

\subsection{Eigenfunctions of the Laplace operator}\

It was shown in \cite{KP,KP2} that both $\,\Sf$- and $\,\Cf$-functions
are eigenfunctions of the same differential operator
\[
L=\left(\alpha_1\partial_1+\alpha_2\partial_2+\dots+\alpha_n\partial_n\right)^2.
\]
Since the matrix of scalar products of simple roots is positive defined, by a suitable choice
of basis, the operator can be brought to the sum of second derivatives with positive coeffinicents,
therefore one may call $\,L\,$ the Laplace operator.

Here we will show that the $\,\E$-functions are eigenfunctions of the $\,L\,$ as well.
We parametrize elements of $\,F^e\,$ by the coordinates in the $\,\omega$-basis
$\,x=\theta_1\omega_1+\dots+\theta_n\omega_n\,$ and denote by $\,\partial_j\,$ the
partial derivative with respect to $\,\theta_j$. Consider the application
of $\,L\,$ to E-functions, we see that they also are its
eigenfunctions:
\[
L\E_\lambda = -4\pi^2\l\lambda\mid\lambda\r\E_\lambda.
\]
In fact, every exponential term in the functions is individually
an eigenfunction of $L$. Since weights of one orbit are
equidistant from the origin, eigenvalues of all terms in each
function coincide. The explicit form of the Laplace operators $\,L\,$
corresponding to the simple Lie group of rank $\,2$ are in
\cite{PZ1}, \cite{PZ2}.

\subsection{Orthogonality and $\,\E$-function transforms}\

Both family of $\,\Cf$- and $\,\E$-function determine a
symmetrized Fourier series expansions. The proof in general for
both families and both continuous and discrete cases is given in
\cite{MP1}. It is based on the orthogonality of $\,\E$-functions
($\,\Cf$-functions) determined by points $\,\lambda\in P_e\,$
(correspondingly $\,\lambda\in P^+\,$). For any
$\,\lambda,\,\lambda'\in P_e\,$ corresponding $\,\E$-functions are
orthogonal on $\,F^e\,$ with respect to Euclidean measure:
\begin{equation}\label{orthog_prop}
\int_{F^e}\E_{\lambda}(x)\overline{\E}_{\lambda'}(x)dx=
|F^e||W_e(\lambda)|\delta_{\lambda\lambda'},
\end{equation}
where bar means a complex conjugation and $\,|F^e|\,$ is a volume
of the fundamental domain $\,F^e\,$. This relation follows from
the orthogonality of the exponential functions for different
weights $\,\lambda\,$ and from the fact that each point $\,\mu\in
P_e\,$ belongs to precisely one $\,W_e$-orbit. Therefore the
$\,\E$-functions corresponding to the points of $\,\lambda\in
P_e\,$ form an orthogonal basis in the Hilbert space of squared
integrable function on $\,F^e$. Therefore, we may expand functions
on $\,F^e\,$ as sums of $\,\E$-functions. Let $\,f\,$ be a
function defined on $\,F^e\,$ then it may be written
\begin{equation}\label{cont_transform}
f(x)=\sum_{\lambda\in\  P^{\,e}} c_{\lambda} \E_{\lambda}(x),
\end{equation}
where $\,c_{\lambda}\,$ is determined by
\begin{equation}
c_{\lambda}=|W_e(\lambda)||F^e|^{-1}\int_{F^e}
f(x)\overline{\E}_{\lambda}(x)dx.
\end{equation}
For details of the proof see \cite{MP1}.

\subsection{Example: $\,\E$-functions
for $\,\Aone\,$}\label{A1section}\

 The $\,\E$-functions of the rank $\,1$ simple Lie group $\,\Aone\,$ happen to be the common exponential functions.
Indeed, the Weyl group of $\,\Aone\,$ has two elements
$\,W=\{id,r\}\,$, where $\,r\,$ is the reflection in the origin of
$\,\R$. The root lattice consists of all even integer points of
$\R$, the weight lattice $P$ is formed by all integers. The even
subgroup $\,W_e\subset W\,$ is the identity element of $W$. Thus
for any point $\,\lambda\,$,  its $\,W_e$-orbit consists of a
single point. Consequently the $\,\E$-function is a single
exponential function.
\begin{equation}\label{a1E}
\E_{\lambda}(x)=\E_m(x)\overset{\text{def}}{=}\sum_{\mu\in\,
W_e(\lambda)}e^{2\pi i\l\mu\mid x\r}=e^{i\pi mx}.
\end{equation}
The fundamental region
$\,F^{\,e}(\Aone)\overset{\text{def}}{=}F(\Aone)\cup
rF(\Aone)=[-1,1]\,$ in $\,\omega$-basis. For this simple case one
can directly verify the decomposition of the products:
\begin{equation}\label{decomp_a_1}
\E_m(x)\overline{\E}_{m^\prime}(x) =\E_{m-m^\prime}(x),\qquad
                \text{for $\ m,\,m^\prime\in\Z$.}
\end{equation}
Consequently we obtain that any two functions $\E_m(x)$,
$\,\E_{m^\prime}(x)$ with $\,m\neq m^\prime\,$, are orthogonal,
i.e.
\begin{equation}\label{orthogA1}
\int_{-1}^1 \E_m(x)\overline{\E}_{m^\prime}(x)dx=
\begin{cases}
0,           &\text{if $m\neq m^\prime$},\\
2,           &\text{if $m=m^\prime$}.
\end{cases}
\end{equation}

The continuous $\,\E$-transform is the expansion
\eqref{cont_transform} of functions over $\,-1\leq x\leq 1\,$
\begin{equation}\label{a1Econt}
f(x)=\sum_{m=-\infty}^{\infty} c_m e^{i\pi mx}, \qquad
c_m=\tfrac12\int_{-1}^1 f(x)e^{-i\pi mx}dx\,.
\end{equation}

\section{Continuous $\E$-transform for simple
Lie groups of rank two}

Next three sections deal with the main topic of the paper, namely
expansions of functions on $F^e$ into  series of $\,\E$-functions
and their inversion (direct and inverse $\,\E$-transforms). In
this section the functions to expand, as well as the
$\,\E$-functions, are continuous ones.  Discrete transforms are
the the subject of Section~5,\,~6.

General structure of the continuous direct and inverse
2-dimensional $\E$-transform is the following:
\begin{equation}\label{continuousET}
\begin{aligned}
f(x,y)&=\sum_{(a,b)\in P_e} c_{a,b}\E_{(a,b)}(x,y)\,,\\
c_{a,b}&=\frac{1}{\int_{F^e}\E_{(a,b)}(x,y)\overline{\E}_{(a,b)}(x,y)\,dx\,dy}
   \int_{F^e}f(x,y)\overline{\E}_{(a,b)}(x,y)\,dx\,dy\,.
\end{aligned}
\end{equation}
Here the overbar indicates complex conjugation.  There are four
pieces of information one needs before the transform
\eqref{continuousET} can be applied to a given function $f(x,y)$.
This information depends on the particular Lie group $G$. We needs
to provide

 (i) \ The infinite set $P_e$ of points of the weight lattice $P$;

 (ii) The finite domain $F^e\in \R^2$;

 (iii) The functions $\,\E_{(a,b)}(x,y)$, $\,(a,b)\in P_e\,$;

 (iv) The normalization coefficients
 $\int_{F^e}\E_{(a,b)}(x,y)\overline{\E}_{(a,b)}(x,y)\,dx\,dy$.
\medskip

There are three compact simple Lie groups of rank two, namely
$\,\SU(3)\,$, $\,\Sp(4)\equiv\O(5)\,$ and $\,\G(2)\,$. Also there
is one semi-simple compact Lie group $\,\SU(2)\times \SU(2)$ which
is not simple. We are using the following notation often used to
denote the corresponding Lie algebras:
$$
\Atwo\leftrightarrow \SU(3)\,,\quad
\Ctwo\leftrightarrow \O(5)\,,\quad
\G_2\leftrightarrow \G(2)\,,\quad
\Aone\times\Aone\leftrightarrow \SU(2)\times\SU(2)\,.
$$

\subsection{The $\,\E$-transforms of $\,\Aone\times\Aone$}\

As we already mentioned in the Introduction there are two ways to
define the even Weyl group for Lie group $\,G=G_1\times G_2\,$
which is a product of two simple groups. More on this subject are
in concluding remarks, Section~9.  Here we give an example when
$\,G=\Aone\times\Aone\,$ and we use $\,W_e(\Aone)\times
W_e(\Aone)\,$ for $\,W_e(\Aone\times\Aone)$. Then this case
becomes a simple concatenation of two cases of $\,\Aone$ described
in \ref{A1section}.

Relative length and angles of the simple roots are given by the scalar products
\[
\l\alpha_1\mid\alpha_2\r=0,\qquad\l\alpha_1\mid\alpha_1\r
=\l\alpha_2\mid\alpha_2\r=2.
\]
Consequently,  $\,\alpha_1=2\omega_1\,$ and
$\,\alpha_2=2\omega_2$. Their dual $\,\check\alpha_k\,$ and
$\,\check\omega_j\,$ coincide with $\,\alpha_k\,$ and
$\,\omega_j$. The root system $\Delta=\{\pm\alpha_1,\pm\alpha_2\}$
geometrically represents the vertices of a square of a side length
$2$. See Figure 1 for the details.

Suppose $\lambda=a\omega_1+b\omega_2$, where $a,b\in\Z$.  Since
$W_e(\Aone)$ consists trivially of its identity element, all
$W_e(\Aone)$-orbits have just one weight, so that,
$$
\E_{(a,b)}(x,y)=e^{\pi i(ax+by)}\,,\qquad a,b\in\Z\,,\quad
x,y\in\R\,.
$$
The fundamental region of $\,W_e(\Aone\times\Aone)\,$ is a direct
product of the fundamental regions of $\,W_e(\Aone)$,
i.e.\begin{equation}\label{fund_reg_a1_a1}
F^e(\Aone\times\Aone)=\{x\omega_1+y\omega_2\mid \text{where
$-1\leq x,\,y\leq 1$}\}.
\end{equation}
Thus we have the familiar extension of the 1-dimensional transform
\eqref{1Dtransform} to two dimensions.

\subsection{The $\,\E$-transforms of $\Ctwo$}\

Relative length and angles of the simple roots of $\Ctwo$ are given
by:
\[
\l\alpha_1\mid\alpha_2\r=-1,\qquad
 \l\alpha_1\mid\alpha_1\r= 1,\qquad
 \l\alpha_2\mid\alpha_2\r= 2\,.
\]
Consequently,
\begin{alignat*}{3}
\alpha_1 &=2\omega_1-\omega_2,
    &\qquad \omega_1&=\alpha_1+\tfrac12\alpha_2,
    &\qquad\check\alpha_1&=2\alpha_1,       \\
\alpha_2 &=-2\omega_2+2\omega_2,
    &\qquad\omega_2&=\alpha_1+\alpha_2,
    &\qquad\check\alpha_2&=\alpha_2.
\end{alignat*}
The root system $\Delta= \{\pm\alpha_1,\pm\alpha_2,\pm(\alpha_1+
\alpha_2),\pm(2\alpha_1+\alpha_2)\}$ geometrically represents the
vertices and midpoints of a square.

\begin{figure}
\includegraphics[width=.7 \textwidth]{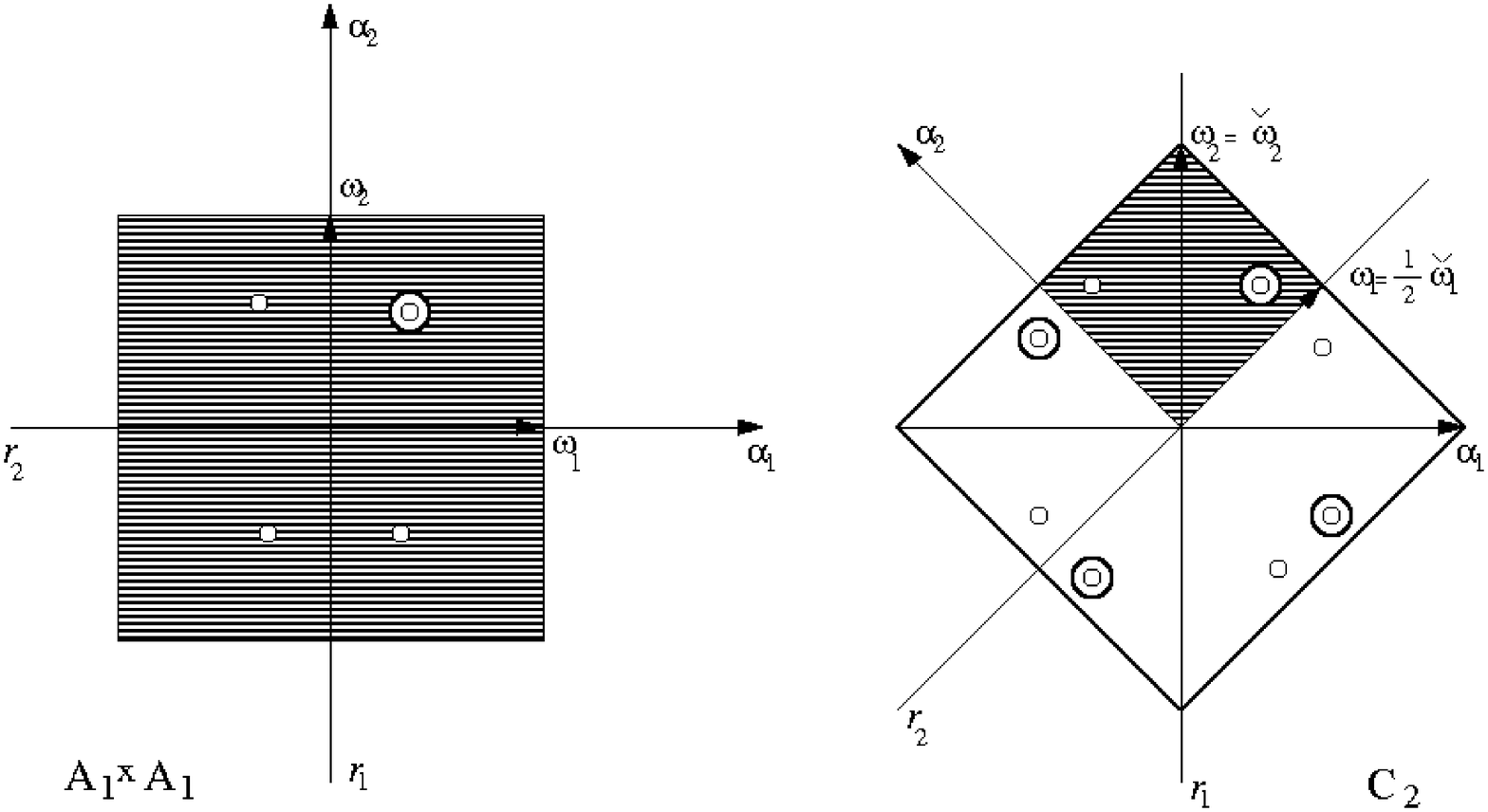}
\caption{\label{fig:roots1} The simple roots, the fundamental
weights, along with their dual, $\,r_1,\, r_2\,$ generators of $\,W$ and
the even fundamental region
(shaded area) for $\Aone\times \Aone$ and $\Ctwo$. The dots $\,\circ\,$ denote
the points of $\,W$-orbit and dots {\bf $\,\bigcirc\,$} denote the points of $\,W_e$-orbit.}
\end{figure}

The fundamental region $F^e(\Ctwo)$ is defined as
$\,F(\Ctwo)\cup r_1 F(\Ctwo)\,$, i.e.
\begin{equation}\label{fund_reg_c2}
F^e(\Ctwo)=\{x\check\omega_1+y\check\omega_2\mid \text{where
$0\leq y\leq 1$ and $0\leq 2x+y\leq 1$}\}.
\end{equation}
Geometrically it is a square with vertices $0$,
$\tfrac{\check\omega_1}2$, $\check\omega_2$ and
$\check\omega_2-\tfrac{\check\omega_1}2$. See Figure~1 for the
details.

We define $\,P_e=P^{+}\cup r_1 P^{++}$. For
$\lambda=(a,b)=a\omega_1+b\omega_2\in P_e$ the even Weyl group
orbit $W_e(\lambda)\equiv W_e(a,b)$ contains either 1 or 4 points:
\[
W_e(a,b)=\begin{cases}
\{(0,0)\} &\text{if $a=b=0$,}\\
\{\pm(a,b), \pm(a+2b,-a-b)\} &\text{if\ $a^2+b^2>0$.}
\end{cases}
\]
According to \eqref{E-function} the $\E$-functions of the Lie
group $\Ctwo$, with $\lambda=a\omega_1+b\omega_2$ and
$z=x\check\omega_1+y\check\omega_2$, are the following
\begin{equation}\label{c2OF}
\begin{array}{l}
\E_{(0,0)}(x,y) =1,\\
\E_{(a,b)}(x,y)
                =2\cos(\pi((2a+2b)x+(a+2b)y))+2\cos(\pi(2bx-ay)),\ \
                      \text{if \ }a^2+b^2>0\,.
\end{array}\end{equation} In particular,
$\,\E_{(a,\,0)}(x,y)=\Cf_{(a,\,0)}(x,y)\,$ and
$\,\E_{(0,\,b)}(x,y)=\Cf_{(\,0,b)}(x,y)$.

\smallskip
To be uniform in both formulas, we introduce a different normalization,
namely, \begin{equation}\label{different_normalization}
\Xi_{(a,b)}(x,y)\define\tfrac{|W_e|}{|W_e(a,b)|}\E_{(a,b)}(x,y),
\end{equation} where $\,|W_e(a,b)|\,$ is the number of points in
$\,W_e(a,b)$. In $\,\Ctwo\,$ case we rewrite \eqref{c2OF} as
$$
\Xi_{(a,b)}(x,y)=2\cos(\pi((2a+2b)x+(a+2b)y))+2\cos(\pi(2bx-ay)),\quad
{\rm for\ all}\ \  (a,b)\in P_e.
$$
For any $\,(a,b),(c,d)\in P_e\,$, orthogonality property is
verified directly,
\begin{gather}\label{c2orthog}
\int_{F^e} \Xi_{(a,b)}(x,y)\overline{\Xi}_{(c,d)}(x,y)dF^e
=\int_0^1dy\int_{-\tfrac y2}^{1-\tfrac y2}
\Xi_{(a,b)}(x,y)\overline{\Xi}_{(c,d)}(x,y)dy\notag\\
=\begin{cases}
0,          &\text{if $a\neq c$ and $b\neq d$,}\\
2,          &\text{if $a=c$ or $b=d$.}
\end{cases}
\end{gather}
In particular, we have $\,\int_{F^e} \Xi_{(a,b)}(x,y)d F^e=0\,$
for any $\,(a,b)\neq(0,0)$.

\begin{figure}
\includegraphics[width=.9 \textwidth]{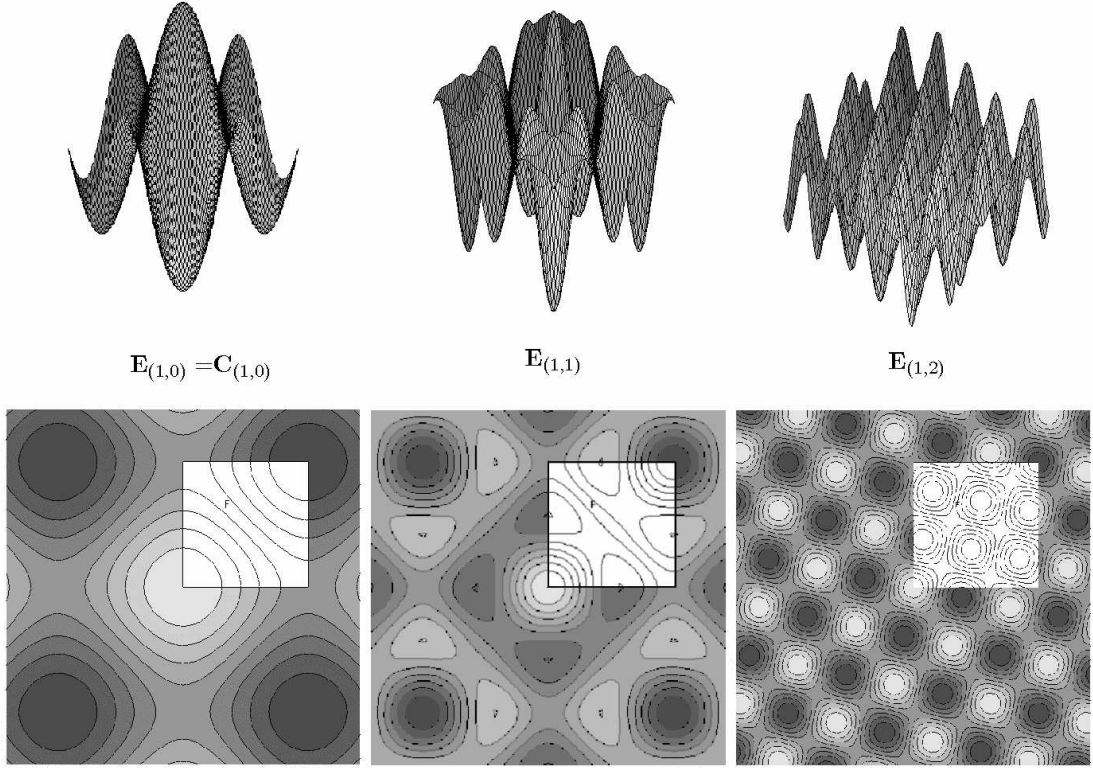}
\caption{\label{fig:roots2} Examples of $\,\E$-functions for $\,\Ctwo.$ }
\end{figure}

\subsection{The $\,\E$-transforms of $\Atwo$}\

Relative length and angles of the simple roots of $\Atwo$ are
given by:
\[
\l\alpha_1\mid\alpha_2\r=-1,\qquad
\l\alpha_1\mid\alpha_1\r=\l\alpha_2\mid\alpha_2\r= 2.
\]
Consequently,
\begin{alignat*}{2}
\alpha_1 &=2\omega_1-\omega_2,
    &\qquad \omega_1&=\tfrac13(2\alpha_1+\alpha_2),      \\
\alpha_2 &=-\omega_1+2\omega_2,
    &\qquad\omega_2&=\tfrac13(\alpha_1+2\alpha_2).
\end{alignat*}
The root system $\Delta= \{\pm\alpha_1,\pm\alpha_2,\pm(\alpha_1+
\alpha_2)\}$ geometrically represents vertices of a regular
hexagon. For details see Figure~3. Take any
$\lambda=a\omega_1+b\omega_2\in P_e=P^+\cup r_1 P^{++}$. Then the
even Weyl group orbit $W_e(a,b)$ contains 1 or 3 points, namely
\[
W_e(a,b)=\begin{cases}
\{(0,0)\} &\text{if $a=b=0$,}\\
\{(a,b), (b,-a-b),(-a-b,a)\} &\text{if\ $a^2+b^2\neq 0.$}\\
\end{cases}
\]
In particular, $\Delta=W_{(1,1)}\cup W_{(-1,-1)}$. Therefore the
$\,\E$-functions of $\Atwo$, with
$\,\lambda=a\omega_1+b\omega_2\,$ and
$\,z=x\check\omega_1+y\check\omega_2\,$, are the following:
\begin{align}
\E_{(0,0)}(x,y) &=1,\notag\\ \E_{(a,b)}(x,y) &=e^{\tfrac{2\pi
         i}3((2a+b)x+(a+2b)y)}+e^{-\tfrac{2\pi i}
         3((x+2y)a+(y-x)b)}+e^{-\tfrac{2\pi i}3((x-y)a+(2x+y)b)}.\notag
\end{align}
In particular,  $\,\E_{(a,0)}(x,y) =  \Cf_{(a,0)}(x,y)\,$ and
$\,\E_{(0,b)}(x,y) =  \Cf_{(0,b)}(x,y)$. Using the normalization
\eqref{different_normalization} we obtain uniform formula
\begin{equation} \Xi_{(a,b)}(x,y) =e^{\tfrac{2\pi
i}3((2a+b)x+(a+2b)y)}+e^{-\tfrac{2\pi i}
3((x+2y)a+(y-x)b)}+e^{-\tfrac{2\pi i}3((x-y)a+(2x+y)b)}.
\end{equation}

\begin{figure}
\includegraphics[width=.7 \textwidth]{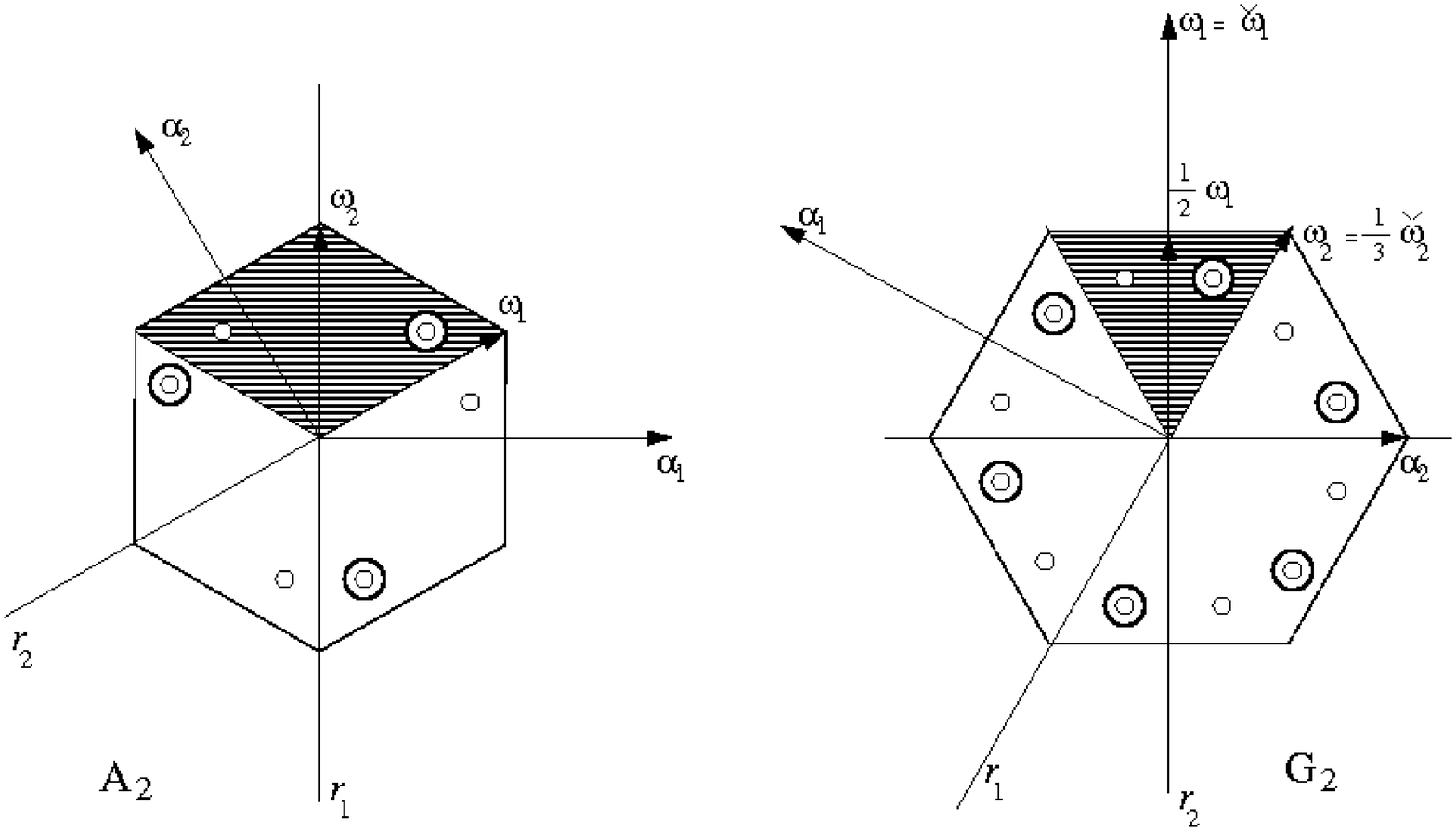}
\caption{\label{fig:roots3} The simple roots, the fundamental
weights, along with their dual, $\,r_1,\, r_2\,$ generators of $\,W$ and
the even fundamental region
(shaded area) for $\Atwo$ and $\Gtwo$. The dots $\,\circ\,$ denote
the points of $\,W$-orbit and $\,\bigcirc\,$ the points of $W_e$-orbit.}
\end{figure}

The fundamental region $F^e(\Atwo)$ is a union of original
fundamental region for Weyl group with its reflection with respect
to $\,r_1\,$, i.e.
\begin{equation}\label{fund_reg_a2}
F^e(\Atwo)=\{x\omega_1+y\omega_2\mid \text{where
$0\leq y\leq 1$ and $0\leq x+y\leq 1$}\}.
\end{equation}
Geometrically it is a rhombus with vertices $0$, $\omega_1$,
$\omega_2$ and $\omega_2-\omega_1$ (see Figure~3).

For any $\,(a,b),(c,d)\in P_e\,$, orthogonality property of $\,\E$-functions of $\Atwo$ can be verified directly,
\begin{gather}\label{a2orthog}
\int_{F^e} \Xi_{(a,b)}(x,y)\overline{\Xi}_{(c,d)}(x,y)d F^e
=\tfrac 1{\sqrt{3}}\int_0^1 dy\int_{-y}^{1-y}
\Xi_{(a,b)}(x,y)\overline{\Xi}_{(c,d)}(x,y)dy\notag\\
=\begin{cases}
0,          &\text{if $a\neq c$ or $b\neq d$,}\\
\sqrt{3},          &\text{if $a=c$ and $b=d$.}
\end{cases}
\end{gather}
\begin{figure}
\includegraphics[width=.9 \textwidth]{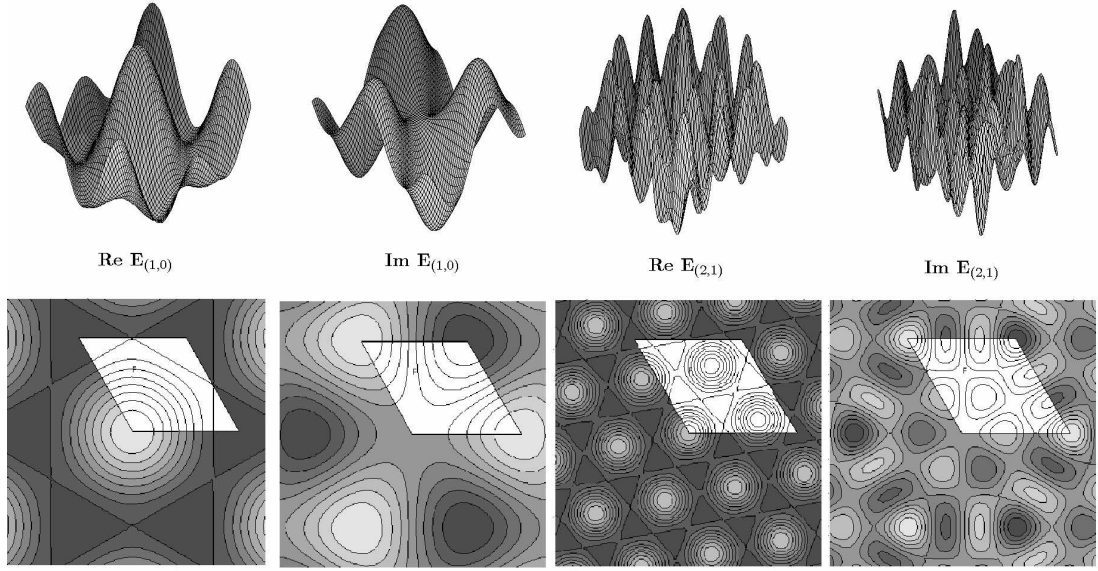}
\caption{\label{fig:roots4} Real and imaginary parts of $\,\E$-functions for $\,\Atwo.$ }
\end{figure}

\subsection{The $\E$-transform of $\Gtwo$}\

Relative length and angles of the simple roots of $\Gtwo$ are
given by:
\[
\l\alpha_1\mid\alpha_2\r=-1,\qquad \l\alpha_1\mid\alpha_1\r=
2,\qquad \l\alpha_2\mid\alpha_2\r= \tfrac23.
\]
Then the relation between simple roots and weights is
\begin{alignat*}{3}
\alpha_1 &=2\omega_1-3\omega_2,
    &\qquad \omega_1&=2\alpha_1+3\alpha_2,
    &\qquad\check\omega_1&=\omega_1,       \\
\alpha_2 &=-\omega_1+2\omega_2,
    &\qquad\omega_2&=\alpha_1+2\alpha_2,
    &\qquad\check\omega_2&=3\omega_2.
\end{alignat*}
There are 12 roots in $\,\Delta(\Gtwo)\,$, namely the following
$$\Delta= \{\pm(2\alpha_1+3\alpha_2),\pm(\alpha_1+3\alpha_2),\pm(\alpha_1+
2\alpha_2),\pm(\alpha_1+\alpha_2),\pm\alpha_1,\pm\alpha_2\}$$
geometrically the roots are vertices of a regular hexagonal star
(see Figure~3).

Let $\lambda=a\omega_1+b\omega_2\in P_e$. Then the even Weyl group
orbit $W_e(\lambda)\equiv W_e(a,b)$ contains 1 or 6 points. More
precisely,
\[
W_e(a,b)=\begin{cases}
\{(0,0)\} &\text{if $a=b=0$,}\\
\{\pm(a,b), \pm(2a+b,-3a-b), \pm(-a-b,3a+2b)\} &\text{if\
$a^2+b^2\neq 0$.}
\end{cases}
\]
The $\,\E$-functions \eqref{different_normalization} of $\Gtwo$,
with $\lambda=a\omega_1+b\omega_2$ and
$z=x\check\omega_1+y\check\omega_2$, are the following:
\begin{align}
\Xi_{(a,b)}(x,y)
&=2\cos(2\pi((2a+b)x+(3a+2b)y))+2\cos(2\pi(ax+(3a+b)y))\\
&  \ +2\cos(2\pi((a+b)x+by)), \quad (a,b)\in P_e.
\end{align}

The fundamental region $F^e(\Gtwo)$ is $\,F(\Gtwo)\cup r_2
F(\Gtwo)\,$:
\[
F^e(\Gtwo)=\{x\check\omega_1+y\check\omega_2\mid \text{where
$0\leq x\leq 1$ and $0\leq 2x+3y\leq 1$}\}.
\]
It is a triangle with vertices $0$, $\tfrac{\check\omega_2}3$ and
$\tfrac{\check\omega_1}2-\tfrac{\check\omega_2}3$ (see Figure~3).
Note that for $\,\Gtwo\,$ we choose the reflection with respect to
$\,r_2$. Therefore we also have to redefine
$\,P^e(\Gtwo)=P^{+}\cup r_2 P^{++}$.

Orthogonality of $\,\E$-functions of $\Gtwo$ can be verified, for
any $\,(a,b),$ $\,(c,d)\in P_e\,$
\begin{gather}\label{g2orthog}
\int_{F^e} \Xi_{(a,b)}(x,y)\overline{\Xi}_{(c,d)}(x,y)d F^e
=\sqrt{3}\int_0^1dx\int_{-\tfrac{2x}3}^{\tfrac{1-2x}3}
\Xi_{(a,b)}(x,y)\overline{\Xi}_{(c,d)}(x,y)dy\notag\\
=\begin{cases}
0,          &\text{if $a\neq c$ or $b\neq d$,}\\
2\sqrt{3},          &\text{if $a=c$ and $b=d$.}\\
\end{cases}
\end{gather}
\begin{figure}
\includegraphics[width=.9 \textwidth]{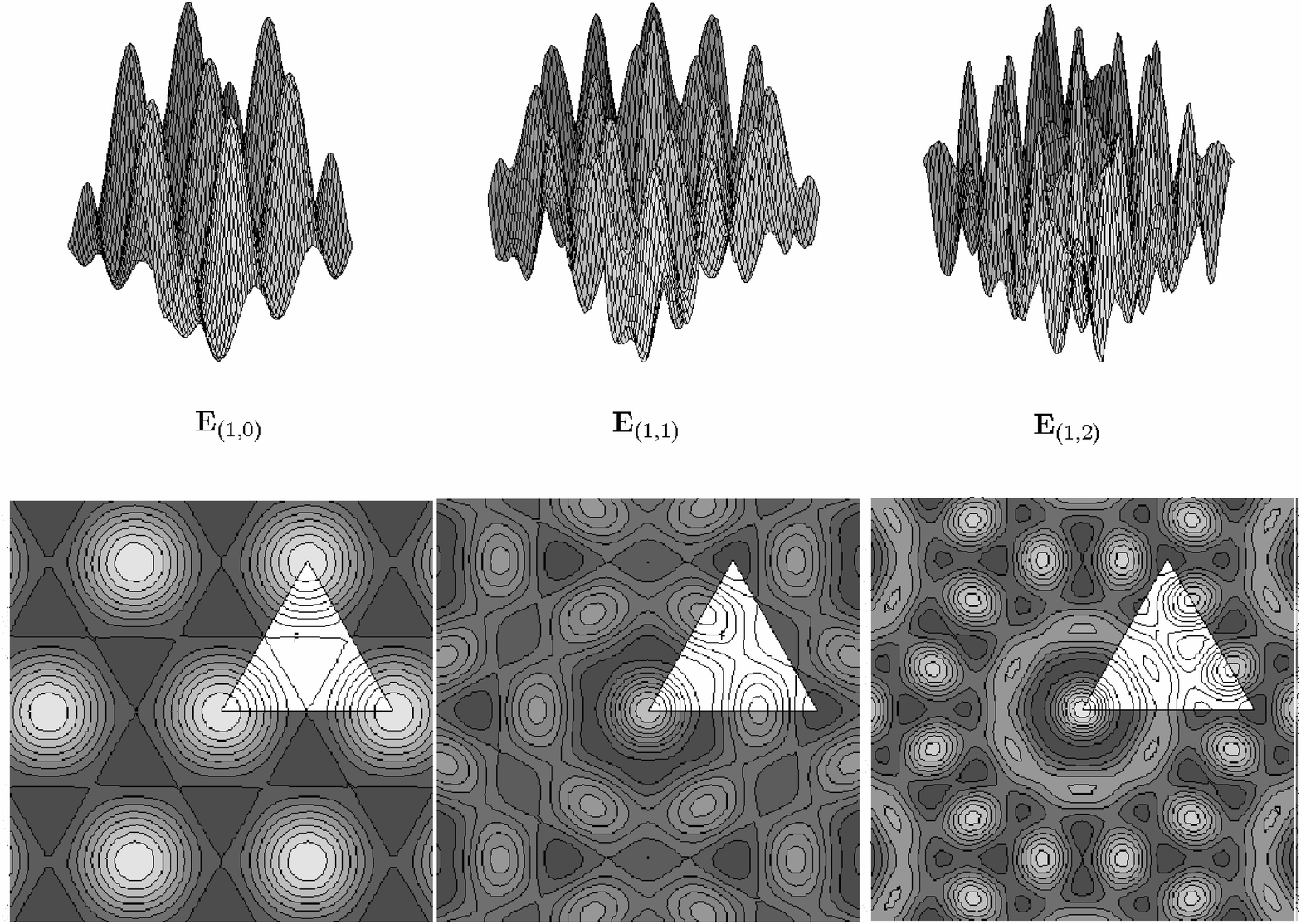}
\caption{\label{fig:roots5} The $\,\E$-functions for $\,\Gtwo.$ }
\end{figure}

\section{A discrete $\,\E$-function transforms}

We introduce the essentials of the discrete finite $\,\E$-function
transform. This transform can be used, for example, to interpolate
values of a function $\,f(x)\,$ between its given values on a
lattice $F^e_M\subset F$. The discretization of $\,\E$-functions
closely parallels that of both $\,\Cf$- and $\,\Sf$-function,
\cite{PZ1}-\cite{PZ3}. All the details of the proof for the
discrete finite $\,\E$-transforms are found in \cite{MP1}. Recall
\eqref{q_lattice}, that the lattice $\,\check Q\,$ is a discrete
$\,W$-invariant subset of $\,\R^n$. Then for any positive integer
$\,M\,$ the set
\begin{equation}\label{hatQ}
T_M=\tfrac1M \check Q/\check Q=\bigcup_{i=1}^n
\tfrac{d_i\check{\alpha}_i}{M}, \quad d_i=0,\dots,M-1
\end{equation}
is finite and $\,W$-invariant. Moreover, $\,T_M\,$
forms the Abelian subgroup of the maximal torus, generated by the
elements of order $M$, of simple compact Lie group corresponding to $\,W$.
One has the basic discrete orthogonal
relation on $\,T_M\,$ for $\,\lambda,\,\mu\in P\,$
\begin{equation}\label{orthogonal_relation}
\sum_{s\in T_M} e^{2\pi\l \lambda, s\r}\overline{e^{2\pi\l \mu, s\r}}=\begin{cases}
|T_M|,          \qquad {\rm if\ \ } \lambda_{|T_M}=\mu_{|T_M},\\
0,          \qquad \text{otherwise.}\\
\end{cases}
\end{equation}

We define the equidistant grid $\,F^e_M\,$ of points in the fundamental region $\,F^e$, namely
\begin{equation}
F^e_M\define F_M\cup
\,r_1(F_M)=\{s=\tfrac{s_1}{M}\check{\omega}_1+\dots+\tfrac{s_n}{M}\check{\omega}_n,\,|
\,s_i\in\Z^{\geq0},\, \sum_{i=1}^n s_im_i\leq M\}\cup \{r_1(s)\,
|\,s\in F_M\,\}
\end{equation}
where $\,m_i$'s are the coefficients of the highest root $\,\xi_h$.

For any two functions $\,f$, $\,g\,$ given by their values on some
$\,F^e_M\,$ we introduce a bilinear form
\begin{equation}\label{herm_form}
\langle f|\,g\rangle_M \define \sum_{s\in F^e_M} \varepsilon_s
f(s)\overline{g(s)}.
\end{equation}
The coefficients $\,\varepsilon_s\,$ in the sum over $F^e_M$ are equal to the
number of points in the torus $\,T_M\,$ that are conjugate to the point
$s\in F^e_M$. By \eqref{orthogonal_relation} for any $\,M\,$ positive integer there exists a finite set
$\,\Lambda_M\,$ of $\,P_e\,$ such that for any
two $\,\lambda,\,\lambda'\in \Lambda_M\,$
\begin{equation}\label{disc_orth}
\langle \Xi_{\lambda}|\,\Xi_{\lambda'}\rangle_M =
\sum_{s\in F^e_M} \varepsilon_s
\Xi_{\lambda}(s)\overline{\Xi(s)}=\delta_{\lambda,\lambda'}\langle
\Xi_{\lambda}|\,\Xi_{\lambda}\rangle_M.
\end{equation}
As a consequence of the orthogonality property \eqref{disc_orth},
we get the following decomposition for any function $\,f(s)\,$
with known values on points of $\,F_M^e$. Indeed, if $\,f(s)\,$ is
given
\begin{equation}\label{equat_disc_transf}
f(s)=\sum_{\Xi_{\lambda}\in\Lambda_M}d_{\lambda}\Xi_{\lambda}(s)
\end{equation}
Then using the orthogonality property \eqref{disc_orth} we may
calculate $\,d_{\lambda}\,$ as
\begin{equation}
d_{\lambda}=\tfrac{\,\langle
f|\,\Xi_{\lambda}\rangle_M}{\langle\,\Xi_{\lambda}|\,\Xi_{\lambda}\rangle_M}
\end{equation}

Once  $\,d_{\lambda}\,$ of the original decomposition
\eqref{equat_disc_transf} were calculated, one can extend discrete
variables in $F_M$ to continues ones:
\begin{equation}
f_{\rm
cont}(x)\define\sum_{\Xi_{\lambda}\in\Lambda_M}d_{\lambda}\Xi_{\lambda}(x).
\end{equation}

It turns out that the function  $\,f_{\rm cont}(x)\,$ smoothly interpolates
the values of $f(s)$, while coinciding with it at the points of $F^e_M$.

Note that to find coefficients $\,\varepsilon_s\,$ one may use the corresponding $\,\Cf$-function coefficients $\,c_s,$
$\,s\in F_M$, which is equal to the number of point in $\,T_M\,$ that are congruent to $\,s$. Indeed, by \eqref{fund_region}
$\,F^e=F\cup r_i F$ for some $\,i\in\{1,\dots, n\}\,$. Then for $\,s=(s_1,\dots,s_n)\in F_M\,$
\begin{equation}\label{rel_c_epsilon}
 \varepsilon_s=\begin{cases}
\tfrac12 c_s,          &\text{if $s_i\neq 0$,}\\
c_s,          &\text{if $s_i=0$.}\\
\end{cases}
\end{equation}

\subsection{Example: discretization of $\Aone$}\

Here we give the description of the $\Aone$ version of the
discrete orthogonality of the $\,\E$-functions.

First, we fix $\,M\in\N\,$, which determines an equidistant grid
of $\,2M+1\,$ points $\,F_M\,$:
\begin{equation}\label{grid_for_e_function}
F^e_M=\{-1,-\tfrac{M-1}{M},\dots,-\tfrac1{M},0,\tfrac1{M},\tfrac2{M},\dots,
\tfrac{M-1}{M},1\}.
\end{equation}

The scalar product in the space of functions defined on
$\,F^e_M\,$ is
\begin{equation}\label{product}
\l f\mid h\r_M\define\sum_{s\in T_M}f(s)h(s)=\sum_{s\in F^e_M}c_s
f(s) h(s).
\end{equation}
For any $\,s\neq\pm 1\,$ there is no other point of $\,T_M\,$ which is conjugated to $\,s\,$ therefore
$\,\varepsilon_s=1$. The point $\,s=1\,$ is conjugate to $\,s=-1\,$  and only one of them belongs to $\,T_M\,$
thus $\,\varepsilon_{-1}=\varepsilon_1=\tfrac12$.
Analogously to the continuous case we obtain the discrete
orthogonality property of the $\,\E$-functions over $F^e_M$:
\begin{equation}\label{a1orthog}
\l \E_m \mid \E_{m^\prime}\r_M=
\begin{cases}
     2M, &\text{if $m=m^\prime \mod 2M$,}\\
     0,  &\text{if $m\neq m^\prime\mod 2M$.}\\
\end{cases}
\end{equation}

Let $f(s)$ be a function with known real values on $F_M$, and be
decomposed as follows,
\begin{equation}\label{decompA1}
f(s)=\sum_{k=-M}^{M} \ d_k \E_k(s),\quad s\in F^e_M.
\end{equation}
Then we can compute the coefficients $d_k$ from
\begin{equation}\label{d_k_coeff}
\l f\mid \E_k\r_M=\sum_{s\in F_M}\varepsilon_s f(s)\E_k(s)=
\begin{cases}
4Md_k, & \text{if $k=-M$ or $k=M$,}\\
2Md_k, & \text{if $k=-M+1,\dots,M-1$.}
\end{cases}
\end{equation}
After the coefficients $\,d_k\,$ have been calculated, one can
replace $\,s\,$ in \eqref{decompA1} by the continuous variable
$\,x\,$:
\begin{equation}\label{CEdecompA1}
f_{\rm cont}(x)\define\sum_{k=-M}^{M} \ d_k
\E_k(x),\quad\text{where $x\in\R$.}
\end{equation}
At $x=s\in F^e_M$,  the continuous function $f_{\rm cont}(x)$
coincides with $f(s)$.

\section{Discretization of two-dimensional transforms}
This section contains all the details of the exploration of the
method of finite $\,\E$-function transform corresponding to the
simple Lie groups of rank $\,2$. General structure of the discrete
$\,2$-dimensional  $\,\E$-transform is the following: for any function $\,f\,$
given on the discrete grid $\,F^e_M\,$
\begin{equation}\label{disc_orth_two_dim}
\begin{aligned}
f_{\rm cont}(x,y)=\sum_{(a,b)\in\,\Lambda_M}d_{(a,b)}\Xi_{(a,b)}(x,y),\\
d_{(a,b)}=\tfrac{\,\langle
f|\,\Xi_{(a,b)}\rangle_M}{\langle\,\Xi_{(a,b)}|\,\Xi_{(a,b)}\rangle_M}. \qquad \qquad \
\end{aligned}
\end{equation}
Here $\,\l\ ,\ \r_M\,$ denotes the Hermitian form
\eqref{herm_form}. For the particular Lie group $\,G\,$ besides
the corresponding $\,\E$-functions there are four other data one
needs to perform transform \eqref{disc_orth_two_dim}:

 (i) \ The finite grid $\,F^e_M \subset F^e$;

 (ii) \ The coefficients $\,\varepsilon_s\,$ for $\,s\in F^e_M\,$;

 (iii) The finite subset $\,\Lambda_M\,$ of $\,P_e$;

 (iv) The normalization coefficients $\,\langle\,\Xi_{(a,b)}|\,\Xi_{(a,b)}\rangle_M$, $\,(a,b)\in \Lambda_M.$

\subsection{Discretization in the case of $\Ctwo$}\

First we describe the grid $\,F^e_M\,$ as in \eqref{grid_for_e_function}. Since the highest root of $\,\Ctwo\,$ is
$\,2\alpha_1+\alpha_2\,$
\[
F^e_M\define\left\{\left(\tfrac{s_1}M,\tfrac{s_2}M\right)\mid
s_0,s_1,s_2\in\Z^{\geq0},\quad
s_0+2s_1+s_2=M>0\right\}\cup\left\{\left(\tfrac{-s_1}M,\tfrac{s_2+2s_1}M\right)\mid
s_2\neq 0\right\}.
\]
See Figure~6 for $\,F^e_M$, $\,M=3,\,4$.

The coefficients $\,c_s\,$ for $\,\Ctwo\,$ are found in \cite{PZ1}. By \eqref{rel_c_epsilon}
\begin{equation}\label{C2coeff}
\varepsilon_s\equiv \varepsilon_{\left(\tfrac{s_1}M,\tfrac{s_2}M\right)}=
\begin{cases}
1, &\text{if $s_1=0$ and $s_2=0,M$,}\\
    &\text{or $s_2=0$ and $s_1=\tfrac M2$,}\\
    &\text{or $s_1=-\tfrac M2$ and $s_2= M$,}\\
2, &\text{if $2s_1+s_2=0$ and $0<s_2<M$,}\\
   &\text{or $2s_1+s_2=M$ and $0<s_2<M$,}\\
   &\text{or $s_2=0$ and $0<s_1<\tfrac M2$,}\\
    &\text{or $s_2=M$ and $-\tfrac M2<s_1<0$,}\\
4    &\text{if $0<s_2<M$ and $0<2s_1+s_2<M$.}
\end{cases}
\end{equation}

The finite set $\,\Lambda_M=\{ (a,b)\in P_e\,|\, 0< a+2b\leq M,\, 0\leq a <M\}$.
Then for any $\,(a,b)\neq (a^\prime,b^\prime)\in\Lambda_M\,$
\[
\l \Xi_{(a,b)}\mid \Xi_{(a^\prime,b^\prime)}\r_M=0,
\]
otherwise, for the set of the lowest pairwise orthogonal
normalized $\,\Xi$-functions:
\[
\l \Xi_{(a,b)}\mid \Xi_{(a,b)}\r_M= 8M^2\times
\begin{cases}
4, &\text{if $\,a=0,\,M\,$ and $\,b=0$,}\\
2, &\text{if $\,a=0\,$ and $\,b=\tfrac M2$,}\\
1, &\text{if $\,0<a<M\,$ and $\,0< a+2b< M$,}\\
   &\text{or $\,a=0\,$ and $\,0< b< \tfrac M2$,}\\
   &\text{or $\,a+2b=M\,$ and $\,0<b<\tfrac M2$}.
\end{cases}
\]
with the higher $\,\Xi$-functions repeating the values of the
lowest ones.

\begin{figure}
\includegraphics[width=.6 \textwidth]{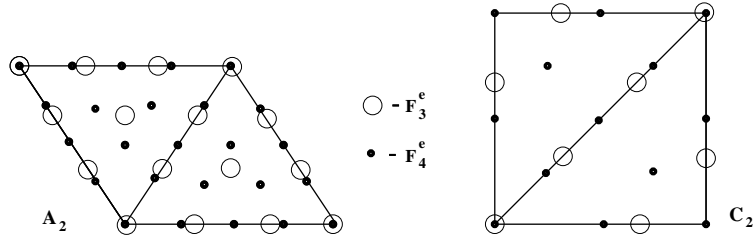}
\caption{\label{fig:roots6} The lattice points of $\,F^e_3,\,F^e_4\,$ in the fundamental region $\,F^e\,$ for $\,\Ctwo\,$ and $\,\Atwo.$}
\end{figure}

\subsection{Discretization in the case of $\Atwo$}\

First we describe the grid $\,F^e_M\,$ as in
\eqref{grid_for_e_function}. The highest root of $\,\Atwo\,$ is
$\,\alpha_1+\alpha_2,$ therefore
\[
F^e_M\define\left\{\left(\tfrac{s_1}M,\tfrac{s_2}M\right)\mid
s_0,s_1,s_2\in\Z^{\geq0},\quad
s_0+s_1+s_2=M>0\right\}\cup\left\{\left(\tfrac{-s_1}M,\tfrac{s_2+s_1}M\right)\mid
s_1\geq0\right\}.
\]
See Figure~6 for $\,F^e_M$, $\,M=3,\,4$.

The coefficients $\,c_s\,$ for $\,\Atwo\,$ are found in
\cite{PZ2}. By \eqref{rel_c_epsilon}
\begin{equation}\label{A2coeff}
\varepsilon_s\equiv \varepsilon_{\left(\tfrac{s_1}M,\tfrac{s_2}M\right)}=
\begin{cases}
\tfrac12 &\text{if $\,s_2=0\,$ and $\,s_1=M$,}\\
         &\text{or $\,s_2=M\,$ and $\,s_1=-M$,}\\
1, &\text{if $\,s_1=0\,$ and $\,s_2=0,M$,}\\
\tfrac32, &\text{if $\,s_2=0\,$ and $\,0<s_1<M$,}\\
    &\text{or $\,s_2=M\,$ and $\,-M<s_1<0$,}\\
   &\text{or $\,s_1+s_2=M\,$ and $\,0<s_2<M$,}\\
   &\text{or $\,s_1+s_2=0\,$ and $\,0<s_2<M$,}\\
3, &\text{if $\,0<s_2<M\,$ and $\,0<s_1+s_2<M$.}
\end{cases}
\end{equation}

The finite set $\,\Lambda_M=\{ (a,b)\in P_e\,|\, 0< a+b\leq M,\,
0\leq a<M\}$. For any $\,(a,b)\neq
(a^\prime,b^\prime)\in\Lambda_M\,$
\[
\l \Xi_{(a,b)}\mid \Xi_{(a^\prime,b^\prime)}\r_M=0,
\]
otherwise, for the set of the lowest pairwise orthogonal
normalized $\Xi$-functions:
\[
\l \Xi_{(a,b)}\mid \Xi_{(a,b)}\r_M= 9M^2\times
\begin{cases}
1, &\text{if $0<a<M$ and $b=0,M$,}\\
   &\text{if $0\leq a+b\leq M$ and $0<a<M$,}\\
3,  &\text{if $a=0$ and $b=0$,}\\
    &\text{or $a=0$ and $b=M$,}\\
    &\text{or $a=M$ and $b=0$.}
\end{cases}
\]
with the higher $\Xi$-functions repeating the values of the
lowest ones.

\subsection{Discretization in the case of $\Gtwo$}\

Since the highest root of $\,\Gtwo\,$ is $\,2\alpha_1+3\alpha_2\,$
the grid $\,F^e_M\,$ is
\[
F^e_M\define\left\{\left(\tfrac{s_1}M,\tfrac{s_2}M\right)\mid
s_0,s_1,s_2\in\Z^{\geq0},\quad s_0+2s_1+3s_2=M>0\right\}\cup
\left\{\left(\tfrac{2s_1+3s_2}M,\tfrac{-s_2}M\right)\mid
s_2\geq0\right\}.
\]

The coefficients $\,c_s\,$ for $\,\Gtwo\,$ are found in \cite{PZ1}. By \eqref{rel_c_epsilon}
\begin{equation}\label{G2coeff}
\varepsilon_s\equiv \varepsilon_{\left(\tfrac{s_1}M,\tfrac{s_2}M\right)}=
\begin{cases}
1, &\text{if $s_1=0$ and $s_2=0$,}\\
    &\text{or $s_1=0$ and $s_2=\tfrac{M}3$,}\\
    &\text{or $s_1=M$ and $s_2=-\tfrac{M}3$,}\\
3, &\text{if $s_1=0$ and $0<s_2<\tfrac{M}3$,}\\
   &\text{or $3s_2+s_1=0$ and $0<s_1<M$,}\\
    &\text{or $3s_2+2s_1=M$ and $0<s_1<M$,}\\
6, &\text{if $0<s_1<M$ and $0<s_1+3s_2$ and $2s_1+3s_2<M$.}
\end{cases}
\end{equation}
The finite set $\,\Lambda_M=\{ (a,b)\in P_e\,|\, 0< 3a+b, \ 3a+2b\leq M,\, 0\leq a\leq \tfrac M3\}$.
Then for any $\,(a,b)\neq (a^\prime,b^\prime)\in\Lambda_M\,$
\[
\l \E_{(a,b)}\mid \E_{(a^\prime,b^\prime)}\r_M=0,
\]
otherwise, for the set of the lowest pairwise orthogonal
normalized $\,\E$-functions:
\[
\l \E_{(a,b)}\mid \E_{(a,b)}\r_M= 6M^2\times
\begin{cases}
6, &\text{if $\,a=0\,$ and $\,b=0$,}\\
3, &\text{if $\,a=\tfrac M3\,$ and $\,b=0$,}\\
2, &\text{if $\,a=0\,$ and $\,b=\tfrac M2$,}\\
1, &\text{or $b=0$ and $0<3a<M$,}\\
   &\text{if $a=0$ and $0<2b<M$,}\\
   &\text{or $0< 3a+b$ and $3a+2b<M$.}\\
\end{cases}
\]
with the higher $\,\E$-functions repeating the values of the
lowest ones.

\section{Decomposition of products of $\,\E$-functions}

For any two points $\,\lambda,\, \lambda'\in P_e\,$ define the
product of corresponding orbits $\,W_e(\lambda)\otimes
W_e(\lambda')\,$ as a set of all points in $\,\R^n\,$ of the form
$\,\mu+\mu'$, $\mu\in W_e(\lambda)$, $\,\mu'\in W_e(\lambda')$.
Since the set of the points $\,\mu+\mu'\,$  $\,\mu\in
W_e(\lambda)$, $\,\mu'\in W_e(\lambda')\,$ is invariant under the
action of corresponding even Weyl group, for any $\,\gamma\in
W_e(\lambda)\otimes W_e(\lambda')\,$ we obtain that
\begin{equation}\label{orb_product}\,W_e(\gamma)\subset
W_e(\lambda)\otimes W_e(\lambda').\end{equation} Therefore any
product of two orbits can be seen as a union of finite number of
orbits of $\,W_e$.

Let both $\,\lambda\,$ and $\,\lambda'\,$ be in $\,P_e\,$ and
\begin{equation}\label{orb_decompos}
W_e(\lambda)\otimes W_e(\lambda')=\bigcup_{\gamma\in\, I}
W_e(\gamma),
\end{equation}
where $\,I\,$ is a finite subset of $\,P_e$. Then for the product
of corresponding $\,\E$-functions we have
\begin{equation}\label{e_func_decomp}
\E_{\lambda}(x)\E_{\lambda'}(x)=\sum_{\gamma\in\,
I}\E_{\gamma}(x).
\end{equation}
Indeed, by \eqref{orb_decompos}
$$
\E_{\lambda}(x)\E_{\lambda'}(x)=\sum_{\mu\in W_e(\lambda)}e^{2\pi
i\langle \mu,\,x\rangle}\sum_{\mu'\in W_e(\lambda')}e^{2\pi
i\langle \mu',\,x\rangle}=\sum_{\gamma\in\, I}\sum_{o(\gamma)\in
W_e(\gamma)}e^{2\pi i\langle
o(\gamma),\,x\rangle}=\sum_{\gamma\in\, I}\E_{\gamma}(x).
$$
For the $\,\E$-functions of $\,\Aone\,$ the product decomposition
was shown in \eqref{decomp_a_1}. However, in higher dimensions of
Euclidean spaces the problem of finding terms of the sum and their
multiplicities in \eqref{e_func_decomp} is a not simple task.
Further in this section we deal with decomposition of the product
of $\,\E$-functions for the three simple Lie groups of rank two.

\subsection{Decomposition of products of $\,\E$-functions for $\,\Ctwo\,$}\

For any $\,(a,b),(c,d)\in P_e\ $ the product in
\eqref{e_func_decomp} can be written as
\begin{equation}\label{prod_e_c2}
\Xi_{(a,b)}\Xi_{(c,d)}
   =\Xi_{(a+c,b+d)}
   +\Xi_{(a-c,b-d)}
   +\Xi_{(a+2d+c,b-c-d)}
   +\Xi_{(a-2d-c,b+d+c)}
   =\sum_{\mu\in W_e(c,d)} \Xi_{(a,b)+\mu}\,
\end{equation}
Here we again we use the normalization
\eqref{different_normalization}. Moreover we can obtain analogous
product decomposition rules for $\,\Cf$-function of the Lie group
$\,\Ctwo\,$. For that we first introduce analogous renormalization
of orbit functions. Namely,
\begin{equation}\label{different_norm_omega}
\Omega_{\lambda}(x)=\frac{|W|}{|W(\lambda)|}\Cf_{\lambda}(x).
\end{equation}
Then from  \eqref{c_via_e} we obtain
\begin{equation}\label{c_via_e_c2}
\begin{aligned}
\Omega_{(a,0)}&=\Xi_{(a,0)},\quad
      \Omega_{(0,b)}=\Xi_{(0,b)}\,,\qquad\,a,b\neq 0\,,\\
\Omega_{(a,b)}&=\Xi_{(a,b)}+\Xi_{(-a,a+b)}.
\end{aligned}
\end{equation}

Combining both \eqref{c_via_e_c2} and \eqref{prod_e_c2} we obtain
the formulas for the product of $\,\Cf$-functions. If $\,(a,b)\in
P^{++}\,$ and $\,(c,d)\in P^{+}\setminus (0,0)\,$
\begin{equation}\label{prod_ome_c2}
\Omega_{(a,b)}\Omega_{(c,d)}=\sum_{\mu\in
W(c,d)}\tfrac{|W|}{|W((a,b)+\mu)|}\Omega_{(a,b)+\mu}.
\end{equation}
The product of two $\,\Cf$-functions determined by points
$\,(a,b)\,$ and $\,(c,d)\,$ is decomposed into the sum of
$\,\Cf$-function labeled by weights from $\,(a,b)+W(c,d)$. In particular,
if both $\,(c,d)\,$ and all the points of $\,(a,b)+W(c,d)\,$ are
in $\,P^{++}\,$, we obtain
\begin{align*}
\Omega_{(a,b)}\Omega_{(c,d)}&=\Omega_{(a+c,b+d)}+\Omega_{(a-c,b-d)}+\Omega_{(-a+c,a+b+d)}+\Omega_{(-a-c,a+b-d)}\\
&+\Omega_{(a+c+2d,b-d)}+\Omega_{(a+c-2d,b+c+d)}+\Omega_{(a+c+2d,b-c-d)}+\Omega_{(a-c-2d,b+d)}.
\end{align*}
The remaining case are
\begin{align*}
\Omega_{(a,0)}\Omega_{(c,0)}=\Omega_{(a+c,0)}+\Omega_{(a-c,0)}+\tfrac{|W|}{|W(a-c,c)|}\Omega_{(a-c,c)}, \qquad \qquad \ \\
\Omega_{(0,b)}\Omega_{(0,d)}=\Omega_{(0,b+d)}+\Omega_{(0,b-d)}+\tfrac{|W|}{|W(2d,b-d)|}\Omega_{(2d,b-d)}, \qquad \qquad \\
\Omega_{(a,0)}\Omega_{(0,d)}=\tfrac{|W|}{|W(a,d)|}\Omega_{(a,d)}+\tfrac{|W|}{|W(a,-d)|}\Omega_{(a,-d)}.
\qquad \qquad
\end{align*}
We also can use these formulas to generalize formulas for tensor
product of $\,W$-orbits of $\,\Ctwo\,$ from section 4.2 in
\cite{KP}.

\subsection{Decomposition of products of $\,\E$-functions for $\,\Atwo$}\

Analogously to the case of $\,\Ctwo\,$ group products of the
$\,\E$-functions of $\Atwo$ decompose into sums of
$\,\E$-functions. For any $\,(a,b),(c,d)\in P_e\,$ we obtain
\begin{equation}\label{prod_e_a2}
\Xi_{(a,b)}\Xi_{(c,d)}=
\Xi_{(a+c,b+d)}+\Xi_{(a+d,b-c-d)}+\Xi_{(a-c-d,b+c)}=\sum_{\mu\in
W_e(c,d)} \Xi_{(a,b)+\mu}.
\end{equation}
The relation \eqref{c_via_e} between $\,\E$- and $\,\Cf$-functions  for $\,\Atwo\,$
is the following
\begin{equation}\label{c_via_e_a2}
\begin{aligned}
\Omega_{(a,0)}=\Xi_{(a,0)},\,\quad \Omega_{(0,b)}=\Xi_{(0,b)}\\
\Omega_{(a,b)}=\Xi_{(a,b)}+\Xi_{(-a,a+b)}.
\end{aligned}
\end{equation}
Here we used the normalization \eqref{different_norm_omega} for
$\,\Cf$-functions. Combining the last two formulas we obtain for
any $\,(a,b)\in P^{++}\,$ and any $\,(c,d)\in P^{+}\setminus
(0,0)\,$
\begin{equation}\label{prod_ome_a2}
\Omega_{(a,b)}\Omega_{(c,d)}=\sum_{\mu\in
W(c,d)}\tfrac{|W|}{|W((a,b)+\mu)|}\Omega_{(a,b)+\mu}.
\end{equation}
The remaining cases:
\begin{align*}
\Omega_{(a,0)}\Omega_{(c,0)}=\Omega_{(a+c,0)}+\tfrac{|W|}{|W(a,-c)|}\Omega_{(a,-c)},
\qquad \qquad\\
\Omega_{(0,b)}\Omega_{(0,d)}=\Omega_{(0,b+d)}+\tfrac{|W|}{|W(-d,b)|}\Omega_{(-d,b)},
\qquad \qquad\\
\Omega_{(a,0)}\Omega_{(0,d)}=\tfrac{|W|}{|W(a,d)|}\Omega_{(a,d)}+\Omega_{(0,-a+d)}.
\qquad \qquad
\end{align*}
Also we have obtained the formulas generalizing formulas for
tensor product of $\,W$-orbits of $\,\Atwo\,$ from section 4.2 in
\cite{KP}.

\subsection{Decomposition of products of $\,\E$-functions for $\,\Gtwo$}\

\noindent The products of the $\,\E$-functions decompose into sums
of $\,\E$-functions. Namely, if $\,(a,b),\,(c,d)\in P_e\,$
\begin{equation}\label{prod_e_g2}
\begin{aligned} \Xi_{(a,b)}\Xi_{(c,d)} = \sum_{\mu\in
W_e(c,d)}\Xi_{(a,b)+\mu(c,d)}=\Xi_{(a+c,b+d)}+\Xi_{(a-c,b-d)}+\Xi_{(a+2d+c,b-3c-d)}\\
+\Xi_{(a-2d-c,b+3c+d)}
+\Xi_{(a-c-d,b+3c+2d)}+\Xi_{(a+c+d,b-3c-2d)}.
\end{aligned}
\end{equation}

In case of $\,\Gtwo\,$ \eqref{c_via_e} gives
\begin{equation}\label{c_via_e_g2}
\begin{aligned}
\Omega_{(a,0)}=\Xi_{(a,0)},\,\qquad \Omega_{(0,b)}=\Xi_{(0,b)}\\
\Omega_{(a,b)}=\Xi_{(a,b)}+\Xi_{(-a,3a+b)}.
\end{aligned}
\end{equation}
As a sequence from \eqref{prod_e_g2} and\eqref{c_via_e_g2} we obtain the decomposition for the product of
$\,\Omega\,$ for $\,(a,b)\in P^{++}\,$ and $\,(c,d)\in P^{+}\setminus (0,0)$
\begin{align} \Omega_{(a,b)}\Omega_{(c,d)}=\sum_{\mu\in
W(c,d)}\tfrac{|W|}{|W((a,b)+\mu)|}\Omega_{(a,b)+\mu}.
\end{align}

\section{Central splitting for $\,\E$-transforms}


The idea of the central splitting of a function $f(x)$ on $F$, or
$F^e$, of a compact semisimple Lie group $G$, is the decomposition
of $f(x)$ into the sum of several component-functions, as many as
is the order $\,s$ of  the center $\,Z\,$ of $G$. Motivation for
considering such splitting is in the property of the
component-functions \cite{MP1}: their $\,\E$-transforms employ
mutually exclusive subsets of $\,\E$-functions of $G$. The
functions $\,\E_\lambda\,$ and $\,\E_{\lambda'}\,$ belong to the
same subset precisely if $\,\lambda\,$ and $\,\lambda'\,$ belong
to the same congruence class.

Let $\,\chi_1,\dots,\chi_s\,$ be the irreducible characters of $\,Z$. Also
any $\,\lambda \in P_e\,$ determines an irreducible character of $\,Z$
\begin{equation}\label{characters}
\chi_{\lambda}\,:\check{z}\mapsto e^{2\pi
i\langle\lambda,\check{z}\rangle}\qquad \check{z}\in Z.
\end{equation}
Then $\,\chi_{\lambda}=\chi_j$  for some $\,1\leq j\leq s$. Then
$\,j\,$ is called the congruence class of $\,\lambda$. It is
constant on the $\,W_e$-orbit of $\,\lambda\,$ and therefore
\begin{equation}
\E_{\lambda}(x+\check{z})=\chi_j(\check{z})\E_{\lambda}(x) \qquad
\check{z}\in Z.
\end{equation}
Thus any $\,f\,$ which is linear combination of $\,\E$-functions
can be written as a sum of $\,s\,$ functions
$\,f(x)=f_1+\dots+f_s$, where
\begin{equation}
f_i(x)=\tfrac1s\sum_{\check{z}\in Z}\overline{\chi_i(\check{z})}
f(x+\check{z}) \qquad 1\leq i\leq s.
\end{equation}

There are two rank-two compact simple Lie groups with non-trivial
center of orders $\,2$ and $\,3$, namely $\,\Ctwo$ and $\,\Atwo$
respectively. The center of $\,\Gtwo\,$ is trivial.

\subsection{Central splitting for $\,\Ctwo\,$}\

As we have already seen in \eqref{fund_reg_c2} the fundamental
region of $\,\Ctwo\,$ is a square with vertices $0$,
$\tfrac{\check\omega_1}2$, $\check\omega_2$ and
$\check\omega_2-\tfrac{\check\omega_1}2.$ The center $\,Z\,$ has
two elements $\,\{0,\check{\omega}_2\}$. According to \cite{MP1}
any function $\,f\,$ on $\,F^e\,$ we may decompose it into
$\,f(x)=f_0(x)+f_1(x)$, where
\begin{equation}\label{decom_c2}
f_0(x)=\tfrac12\{f(x)+f(x+\check{\omega}_2)\} \quad {\rm and}\quad
f_1(x)=\tfrac12\{f(x)-f(x+\check{\omega}_2)\}.\end{equation}
However for any $\,x=a\check{\omega}_1+b\check{\omega}_2\in F^e\,$
the point $\,x+\check{\omega}_2\,$ is outside of $\,F^e$. By
suitable transformation we bring it back to the fundamental
region:
$$r_{\alpha_2,1}r_{\alpha_1}r_{\alpha_2,1}r_{\alpha_1}(a,b+1)=(-a,1-b).$$
Therefore the component function can be written for $\,x\,$ in
$\,\check{\omega}$-basis as
$$\begin{array}{c}
f_0(a,b)=\tfrac12\{f(a,b)+f(-a,1-b)\}\\
f_1(a,b)=\tfrac12\{f(a,b)-f(-a,1-b)\}.
\end{array}
$$

The main property of both $\,f_0\,$ and $\,f_1\,$ is that each of
them decomposes into a linear combination of $\,\E$-functions from
one congruence class only $\,0\,$ for $\,f_0\,$ and $\,1\,$ for
$\,f_1$.
\subsection{Central splitting for $\,\Atwo\,$}\

In the case of $\,\Atwo\,$ the fundamental region is a rhombus
with vertices $0$, $\omega_1$, $\omega_2$ and
$\omega_2-\omega_1$ as in \eqref{fund_reg_a2}. The center
$\,Z\,$ has three elements
$\,\{0,{\omega}_1,{\omega}_2\}\,$ and any function on
$\,F^e\,$ is decomposed into the sum of three function
$\,f(x)=f_0(x)+f_1(x)+f_2(x)\,$ where
$$
\begin{array}{l}
f_0(x)=\tfrac13\{f(x)+f(x+{\omega}_1)+f(x+{\omega}_2)\},\\
f_1(x)=\tfrac13\{f(x)+ e^{-2\pi i/3}f(x+{\omega}_1)+e^{-2\pi i/3}f(x+{\omega}_2)\},\\
f_2(x)=\tfrac13\{f(x)+ e^{2\pi i/3}f(x+{\omega}_1)+e^{2\pi
i/3}f(x+{\omega}_2)\}.
\end{array}
$$
Again for $\,x=a{\omega}_1+b{\omega}_2\in F^e\,$ both
$\,x+{\omega}_1\,$ and $\,x+{\omega}_2\,$ are not necessarily in
$\,F^e$. We have $\,x+{\omega}_1=(a+1,b)\,$ and
$\,x+{\omega}_2=(a,b+1)$. Then there are two cases for points
$\,(a,b+1)\,$ and $\,(a+1,b)\,$ to be brought to $\,F^e$ by
suitable transformations from the affine Weyl group of $\,\Atwo$.
$$\begin{array}{c}
r_{\alpha_2,1}r_{\alpha_1}(a,b+1)= (b,-a-b+1)\\
r_{\alpha_1,1}r_{\alpha_2}(a+1,b)=(1-a-b,a)\end{array} \quad {\rm
for} \ a\geq 0$$ and
$$\begin{array}{c}
r_{\alpha_1}r_{\alpha_2,1}(a,b+1)= (b-1,a+1)\\
r_{\alpha_2,1}r_{\alpha_1}(a+1,b)=(b-1,-a-b+1)\end{array} \quad
{\rm for} \ a< 0.$$

Finally, one obtains, for $\,x=(a,b)\in F^e$, $\,a\geq 0$
$$
\begin{array}{c}
f_0(a,b)=\tfrac13\{f(a,b)+f(b-1,-a-b+1)+f(b-1,a+1)\}\\
f_1(a,b)=\tfrac13\{f(a,b)+ e^{-2\pi i/3}f(b-1,-a-b+1)+e^{-2\pi i/3}f(b-1,a+1)\},\\
f_2(a,b)=\tfrac13\{f(a,b)+ e^{2\pi i/3}f(b-1,-a-b+1)+e^{2\pi
i/3}f(b-1,a+1)\}
\end{array}
$$
or if $\,x=(a,b)\in F^e$, $\,a<0\,$
$$
\begin{array}{c}
f_0(a,b)=\tfrac13\{f(a,b)+f(1-a-b,a)+f(b,-a-b+1)\}\\
f_1(a,b)=\tfrac13\{f(a,b)+ e^{-2\pi i/3}f(1-a-b,a)+e^{-2\pi i/3}f(b,-a-b+1)\},\\
f_2(a,b)=\tfrac13\{f(a,b)+ e^{2\pi i/3}f(1-a-b,a)+e^{2\pi
i/3}f(b,-a-b+1)\}.
\end{array}
$$

Again each $\,f_i,$ $\,i=0,1,2\,$ in this sum may be decomposed as
a sum of $\,\E$-functions from the congruence class $\,i$.

\section{Concluding remarks}

1.\ Similarly as the $\Cf$- and $\Sf$-functions, the
$\,\E$-functions can be viewed as a family of orthogonal
polynomials, related to a particular semisimple Lie group and to a
particular $W_e$-orbit, in as many variables as is the rank of the
group. Such variables, as we use them here, are constrained to the
$n$-dimensional torus of the appropriate Lie group. The
polynomials have many properties of traditional special functions.
Easy discretization of the polynomials is an unusual feature,
particularly in a multidimensional set up.

\smallskip
2. \ The $\,\E$-functions of $\,\SU(2)\,$ are common exponential
function in one variable
$$
\E_m(x)=e^{imx}=\tfrac12(\Cf_m(x)+\Sf_m(x)).
$$
Roughly speaking, $\,\E$-function are related to $\,\Cf$- and
$\,\Sf$-functions as the exponential function is related to cosine
and sine functions. The special role of imaginary unit does not
seem to generalize.

\smallskip
3. \ Besides three introduced transforms in the case when $\,G\,$
is semisimple there are other derived transforms which may be
considered. Suppose $\,G=G_1\times G_2$, where $\,G_1\,$ and
$\,G_2$. Let $\,F_1\,$ be either the fundamental region of
$\,G_1\,$ or fundamental region of its even subgroup and $\,F_2\,$
respectively for $\,G_2$. Then any function $\,f(x_1,x_2)\,$ on
$\,F_1\times F_2\,$ can be expanded using any of $\,\Cf$-
$\,\Sf$-, $\,\E$-functions on $\,F_1\,$ and any of these three
types on $\,F_2$. Thus one may have $\,\E\Cf$ or
$\,\Sf\E$-transforms rather then $\,\E\E$-transforms we studied in
the main body of the paper.

\smallskip
4. \ The $\E$-functions are complex valued, in general. A function $\E_\lambda(x)$ is
real precisely if the orbit $W_e(\lambda)$ contains both weights $\pm\lambda$. More about
when that happens see in \cite{MP1}.

\smallskip
5. \  A choice of $\,W_e^{\rm aff}$-fundamental domain  $F^e$ is
not unique. It is made out of two adjacent copies of the
fundamental domain $F$ of $W$. One can flip $F$ in any of its
$(n-1)$\-dimensional faces in order to get $F^e$. Obviously, for
any choice of $\,F^e\,$ one can introduce both continuous and
discrete transform: general theory allows one to set up  $\,P_e$,
$\,F^e_M$ etc. It is conceivable that practical considerations may
dictate preferred choice.

\smallskip
6. \ As we already mentioned in the introduction there are two ways to define
even Weyl group in case when original Lie group is a product of two simple Lie groups
$\,G=G_1\times G_2$. First possibility is when
$\E$-transform of $G$ is taken up to be the simultaneous
$\E$-transform of $G_1$ and the $\E$-transform of $G_2$. The
$\E$-functions of $G$ are products of $\E$-functions of $G_1$ and
the $\E$-functions of $G_2$. In this case  we take $W_e(G)$ to be
$W_e(G_1)\times W_e(G_2)$.

The second possibility arises from the fact that
\begin{equation}\label{different_way_on_product}W_e(G_1\times G_2)\neq W_e(G_1)\times W_e(G_2).\end{equation}
Consider for example $G=\SU(2)\times \SU(2)$. Its Weyl group is of
order four, its elements being $1,\ r_1,\ r_2,\ r_1r_2$. Hence
$W_e(\SU(2)\times \SU(2))$ has two elements, namely $1$ and
$r_1r_2$. In three dimensions the possibilities this option allows
are more curious, interesting and involved. We are going to pursue
them elsewhere.

\smallskip
7. \  The most important qualitative argument in favor of
efficiency of discrete and continuous expansions of functions
given on $F$ into series of either $\Cf$- or $\Sf$- or
$\E$-functions, is that they involve discrete groups larger than
the translation group of traditional Fourier expansions. Indeed,
it is the affine Weyl group acting in $\R^n$, which contains the
translations as its subgroup. More specifically, the fundamental
region of the translation group is the proximity cell (Voronoi
domain) $V$ of the root lattice of $G$, while the fundamental
region $F$ for the affine Weyl group is much smaller,
$|V|=|F||W|$. Thus the larger is is the order $|W|$ of the Weyl
group, the more efficient are our expansions (fewer `harmonics'
needed). The Voronoi domains of root lattices for all simple $G$
are described in \cite{MP3}.

Independently interesting would be to study multidimensional
Fourier expansions in general, that is expansions based on
translational symmetry, as opposed the reflection symmetry of the
affine Weyl group we use. In that case Voronoi domains would play
the role of $F$ here, because they are the tiles filling the space
by translations. Suppose that one wants to insists on expansions
based on translation symmetries like $\,x\mapsto x+2\pi\,$ in
$\,1$-dimension. Then the corresponding symmetry group is the
translation subgroup of the affine Weyl group $\,W^{\aff}$. The
expansions then refer to functions given on the fundamental region
of $\,W^{\aff}\,$ which is proximity cell (Voronoi domain) of the
root lattice of $\,G$. Translations then tile the entire
$\,n$-dimensional space by copies of the proximity cell. A
description of the cells for all simple Lie groups is found in
\cite{KP}.

\smallskip
8. \ Finally let us point out several questions naturally arising
from this work and its possible extensions. There are two
$\E$-transforms on square lattices of $\R^2$ related to the groups
$\SU(2)\times \SU(2)$ and $\O(5)$. How do they compare? Similarly
there is $\E$-transforms on triangular lattices of
$\G(2)$ and $\Cf$-transform on the same lattice of $\SU(3)$. When
to use one and when the other? Such
dilemmas grow rapidly with the dimension of the transform. Thus in
3D there are four  $\E$-transforms on cubic lattices.

To know more about computing efficiency of the transforms would be very useful.

Restriction of the Lie group $G$ to, say, its maximal reductive
subgroup $G'$ implies reduction of the $\,\E$-functions of $G$ to
the sum of $\,E$-functions of $G'$. Calculate such branching
rules.

There are finitely few discrete points in $F$ (for each semisimple
$G$) where all $\,\Cf$-functions take integer values. Are there
points with this property also for  $\E$-functions? A trivially
affirmative answer is given by $\E_\lambda(0)$ for all $\lambda$
and all $G$.

Symmetrization and antisymmetrizarion of tensor powers of $W_e$
orbits results in the sum of several orbits. In terms of
$\E$-functions such an uncommon multiplication would yield a sum
of $\E$-functions. Any use for it?

\section{Acknowledgments}
We are grateful for the partial support for this work from the National Science and
Engineering Research Council of Canada, MITACS, the MIND Institute of Costa Mesa,
California, and to Lockheed Martin Canada. We are
also grateful to J.P. Gazeau and A. Klimyk for their
helpful comments and to A. Zaratsyan for preparing the early version of the figures
in the paper. One of the authors (IK) acknowledges the hospitality of the
Centre de recherches math\'ematiques,
Universit\'e de Montr\'eal.


\end{document}